\pdfoutput=1
\documentclass[11pt]{article}
\usepackage{graphicx}
\usepackage{slashbox}
\usepackage{multirow}
\usepackage{bm}
\usepackage{cite}
\usepackage{changepage}
\usepackage{psfig}
\usepackage{subfigure}

\catcode`@=12
\makeatletter
\@addtoreset{equation}{section}

\makeatother
\begin{document}
\title{\LARGE \bf Flavored leptogenesis with quasi degenerate neutrinos in a 
broken cyclic symmetric model}
\author{{\bf Biswajit Adhikary$^{\rm a}$\footnote{biswajitadhikary@gmail.com}, Mainak Chakraborty$^{\rm b}$\footnote{mainak.chakraborty@saha.ac.in}, 
Ambar Ghosal$^{\rm b}$\footnote{ambar.ghosal@saha.ac.in}}\\
a)Department of Physics, Gurudas College,
Narkeldanga,
 Kolkata-700054, India\\
b) Saha Institute of Nuclear Physics, 1/AF Bidhannagar,
  Kolkata 700064, India 
  }
\maketitle
\begin{abstract}
 Cyclic symmetry in the neutrino sector with the type-I seesaw mechanism in the mass basis of charged leptons
and right chiral neutrinos ($N_{iR}$, $i=e,\mu,\tau$) generates two fold degenerate light neutrino and three 
fold degenerate heavy  neutrino mass spectrum. Consequently, such scheme, produces vanishing one light neutrino mass 
squared difference and lepton asymmetry. To circumvent such unphysical outcome, we break cyclic symmetry
in the diagonal right chiral neutrino mass term by a small breaking parameter.
~Nonzero mass squared differences and mixing angles are generated with the help of the small breaking parameter.
Smallness of the breaking parameter opens up a possibility of resonant leptogenesis. Assuming complex Yukawa 
couplings, we derive generalized expressions flavor dependent CP asymmetry parameters ($\varepsilon^\alpha_i$)
which are valid for quasi degenerate as well as hierarchical mass spectrum of right handed neutrinos.
 There after we set up the chain of coupled Boltzmann equations
(which are flavor dependent too) which have to be solved in order to get the final lepton asymmetries.
Depending upon the temperature regime the CP asymmetries and the Boltzmann equations may also be flavor
independent. As our goal is to study the enhancement of CP asymmetry due to quasi degeneracy of right handed 
neutrinos, we select only the lowest allowed (by neutrino oscillation data) value of breaking parameter
(and other corresponding Lagrangian parameters) and estimate the baryon asymmetry parameter $Y_B$.
Experimental constraint of $Y_B$ introduces a bound on right handed neutrino mass 
 which remained unrestricted by neutrino oscillation data.

\end{abstract}
\newpage
\section{Introduction}
Many experimental observations suggest the excess of matter over antimatter in the universe.
In fact, no evidence of appreciable amount of antimatter has been found yet. 
Various considerations indicate that the universe has started
its evolution from a baryon symmetric state and the baryon asymmetry observed in the present 
era is generated dynamically. The process responsible for the generation of baryon asymmetry
is known as Baryogenesis\cite{Rubakov:1996vz,Trodden:1998ym,Riotto:1998bt,Cline:2006ts,Dine:2003ax}. 
There are three necessary conditions known as Sakharov conditions\cite{sak}
which have to be satisfied in order to generate baryon asymmetry dynamically. They are
(i)Baryon number violation, (ii)C and CP violation, (iii)departure from thermal equilibrium.
The baryon asymmetry of the universe is expressed popularly by two nearly equivalent parameters 
$\eta_B$ and $Y_B$, mathematically which can be written as  
\begin{eqnarray}
&&\eta_B=\frac{n_B-n_{\overline{B}}}{n_\gamma} \\
&&Y_B=\frac{n_B-n_{\overline{B}}}{s}
\end{eqnarray}
where $n_B,n_{\overline{B}},n_\gamma$ are number densities of baryons, antibaryons, photons respectively and $s$ is the entropy density.
After the recent result of Planck satellite experiment, the value of $\eta_B$\footnote{The value of $\eta_B$ and $Y_B$ at present
epoch are related as $\eta_B=7.04 Y_B$.} can vary mostly within the range
as $(6.02 - 6.18)\times10^{-10}$\cite{Dev:2015uca,Ade:2015xua,Ade:2013zuv}. The lower limit arises solely due to the analysis of the Planck data at $68\%$
limit whereas inclusion of gravitational lensing data with the above shifts the value of $\eta_B$ to the higher end.
\paragraph{}
Among the various existing mechanisms to generate baryon asymmetry at electroweak scale, baryogenesis via 
leptogenesis\cite{one, one2, khlopov, Davidson:2008bu}
is a simple and attractive mechanism. In this mechanism lepton asymmetry generated at a high scale ($\sim 10^9$ Gev) gets
converted into baryon asymmetry ($\eta_B$) at electroweak scale due to $B+L$ violating sphaleron interactions\cite{Buchmuller:2000as,Bernreuther:2002uj}. Within
the framework of Standard $SU(2)_L\times U(1)_Y$ Model (SM) with at least two right chiral neutrinos ($N_{iR}$) there is a Dirac type
Yukawa interaction of $N_{iR}$ with electroweak leptons and Higgs doublet. At a high scale where 
$SU(2)_L\times U(1)_Y$ is unbroken, $N_{iR}$'s with definite mass can decay into both (i) charged lepton
with charged scalar $(N_{iR} \rightarrow e^-_\alpha,\phi^+)$ and (ii) light neutrino with neutral scalar 
$(N_{iR} \rightarrow \nu_\alpha, \phi^0)$. 
CP conjugate decays of the above processes are also admitted due to Majorana property of $N_{iR}$.
If out of equilibrium decay of $N_{iR}$ in conjugate process occurs at different rate than the actual 
process, a net lepton number asymmetry will be realized, which then gets converted into baryon asymmetry due to
sphaleronic interactions of the SM.
\paragraph{}
In the present work, we investigate the interrelation between leptogenesis, heavy right chiral neutrinos and flavor mixing
of light neutrinos. In fact, we first constrain the parameter space utilizing extant neutrino oscillation 
data\cite{Forero:2014bxa, GonzalezGarcia:2012sz, Tortola:2012te},
and subsequently we further restrict the parameter space incorporating the reported value of baryon asymmetry by Planck satellite experiment. 
In particular, we consider a well defined model based on $SU(2)_L\times U(1)_Y$ gauge symmetry 
with three right chiral neutrinos $N_{e_R}$, $N_{\mu_R}$, $N_{\tau_R}$ invoking type-I seesaw mechanism and discrete 
cyclic symmetry. The model has already been investigated by the authors recently in the context of an application of the 
general methodology developed to calculate three mass eigenvalues, three mixing angles and Dirac and Majorana phases 
of a general complex $3\times3$ Majorana neutrino mass matrix. In this work, we study baryogenesis via leptogenesis 
in detail.
\paragraph{}
We briefly describe the model here. The cyclic symmetry considered as follows
\begin{eqnarray}
\nu_{e_L}\rightarrow \nu_{\mu_L} \rightarrow \nu_{\tau_L} 
\rightarrow \nu_{e_L},\nonumber\\
N_{e_R}\rightarrow N_{\mu_R} \rightarrow N_{\tau_R} \rightarrow N_{e_R}.\label{cyclic}
\end{eqnarray}
The symmetry invariant neutrino mass matrix can generate nonzero $\theta_{13}$ and other two mixing angles within the experimentally
constrained range at the leading order. In spite of having those attractive properties the effective neutrino mass matrix 
encounters a serious problem of degenerate eigenvalues which is strictly forbidden by the neutrino oscillation experimental data. 
Due to such degeneracy in eigenvalues the mixing angles can not be determined uniquely. To overcome those shortcomings the cyclic
symmetry is broken in the right chiral neutrino sector only and the effective neutrino mass matrix is constructed again
with this broken symmetric right handed neutrino mass matrix and symmetry preserving Dirac neutrino mass matrix.
The eigenvalues and mixing angles of this broken symmetric effective neutrino mass matrix are calculated directly (without
any perturbative approach) using the generalized formulas\cite{Adhikary:2013bma}.
\paragraph{}
The second part of the work deals with generation of baryon asymmetry through the production of lepton asymmetry.
In our symmetry breaking scheme the breaking parameter is taken to be small and hence the masses of the three
right handed neutrinos are not far apart from each other. Therefore, instead of hierarchical leptogenesis here
we have to use the resonant leptogenesis formalism. Again we know that the lepton flavors $(e,~\mu,~\tau)$ involved 
in the process may or may not be separately distinguishable depending upon the temperature regime in 
which we are working, therefore, the study of leptogenesis is done in three different regimes, viz
{\bf (i) fully flavored $(m({\rm GeV})<10^9)$:} three lepton flavors $(e,~\mu,~\tau)$ are completely distinguishable,
{\bf (ii) $\tau$-flavored $(10^9<m({\rm GeV})<10^{12})$:} we can't differentiate between $e$ and $\mu$ but $\tau$ is distinguishable,
{\bf (iii) unflavored $(m({\rm GeV})>10^{12})$:} all three flavors act indistinguishably. At first the expressions
of flavor dependent CP asymmetry parameters are obtained for resonant leptogenesis formalism (CP asymmetry 
parameters required for the other two cases ((ii) and (iii)) can be obtained by summing over the flavor indices).
These CP asymmetry parameters are then inserted into Boltzmann equations which have to be solved to get the final value of the lepton 
asymmetry. This lepton asymmetry will be converted into baryon asymmetry through sphaleron process.
The CP asymmetry parameters and several decay and scattering terms of the Boltzmann equation involve 
Lagrangian parameters which are already constrained by neutrino oscillation data. Our parametrization 
of the neutrino mass matrix is such that the right handed neutrino mass remains unrestricted by the 
oscillation data. Calculation of baryon asymmetry parameter ($\eta_B ~{\rm or}~Y_B$) requires mass of the right 
handed neutrino along with other restricted set of parameters including the phases. The experimental bound on $\eta_B$ 
introduces a limit on the mass of the right handed neutrino and the signs of the phase parameters also get fixed.

\paragraph{}
We organize the present work as follows: In Section \ref{mod} we briefly discuss the model under consideration. Starting from a most 
general leptonic mass term we have generated the effective neutrino mass matrix($m_\nu$) through type-I seesaw mechanism. 
Parametrization and diagonalization of the broken symmetric mass matrix is also described in brief in this section.
Different subsections of Section \ref{baryo} deals with the detailed mathematical expressions of flavor dependent 
CP asymmetry parameters and chain of coupled Boltzmann equations which are solved to obtain the flavor dependent/independent 
lepton asymmetry. Section \ref{baryo1} contains the recipe to get the baryon asymmetry from lepton asymmetry in 
different energy regimes. Outcome of the numerical analysis for various cases are presented in Section \ref{nu_ph}. 
Finally we summarize the whole analysis in Section \ref{summ}. 
\section{Cyclic symmetric model}\label{mod}
The most general leptonic Yukawa terms of the Lagrangian in the present model is 
\begin{equation}
-\mathcal{L}_{\rm mass}=(m_{\ell})_{ll'} \overline{l_L} l'_R + m_{D_{ll'}}\overline{\nu_{lL}} N_{l'R} + 
M_{R_{ll'}} \overline{N^c_{lL}} N_{l'R}
\end{equation}
where $l,~l'=e,~\mu,~\tau$.
We demand that the neutrino part of the Lagrangian is invariant under the cyclic permutation 
\cite{Adhikary:2013bma,Koide:2000zi,Damanik:2007cs,Damanik:2010rv,Samanta:2015oqa} as given in eq.(\ref{cyclic}).
The cyclic symmetric Dirac neutrino mass matrix $m_D$ takes the form
\begin{equation}
m_D=\left( \begin{array}{ccc}
     y_1 & y_2 & y_3\\ y_3 & y_1 & y_2 \\ y_2 & y_3& y_1 \\
    \end{array}\right)  \label{md}
\end{equation}
where in general all the entries are complex. The matrix $m_D$ can be written in terms of Yukawa couplings as
$(m_D)_{ij}=h^\nu_{ij}\frac{v}{\sqrt{2}}$, where $h^\nu_{ij}$ are the Yukawa couplings and $v$ is the VEV ($v=246~{\rm GeV}$).
We assume a  basis in which the right 
handed neutrino mass matrix ($M_R$) and  charged lepton mass matrix ($m_\ell$) are mass
diagonal. Further, imposition of cyclic symmetry dictates the texture of $M_R$ as 
\begin{equation}
M_R=\left( \begin{array}{ccc}
     m & 0 & 0\\0 & m& 0\\0& 0& m\\ 
    \end{array}\right).
\label{mr}
\end{equation}
Invoking type-I seesaw mechanism the 
effective neutrino mass matrix $m_\nu$,  
\begin{equation}
 m_\nu=-m_D M_R^{-1} m_D^{T}
\end{equation}
takes the form as 
\begin{equation}
m_\nu=-\frac{1}{m}\left( \begin{array}{ccc}
                   y_1^2+y_2^2+y_3^2 & y_1 y_2+y_2 y_3+y_3 y_1 &y_1 y_2+y_2 y_3+y_3 y_1 \\
                  y_1 y_2+y_2 y_3+y_3 y_1 & y_1^2+y_2^2+y_3^2 & y_1 y_2+y_2 y_3+y_3 y_1\\
                  y_1 y_2+y_2 y_3+y_3 y_1 &y_1 y_2+y_2 y_3+y_3 y_1 & y_1^2+y_2^2+y_3^2 \\
                  \end{array}\right).
\label{effm}
\end{equation}
With a suitable choice of parameters $m_\nu$ can be rewritten as 
\begin{equation}
m_\nu=m_0 \left( \begin{array}{ccc}
1+p^2e^{2i\alpha}+q^2e^{2i\beta} & pe^{i\alpha}+qe^{i\beta}+pqe^{i(\alpha+\beta)}& 
pe^{i\alpha}+qe^{i\beta}+pqe^{i(\alpha+\beta)}\\
pe^{i\alpha}+qe^{i\beta}+pqe^{i(\alpha+\beta)} & 1+p^2e^{2i\alpha}+q^2e^{2i\beta} & pe^{i\alpha}
+qe^{i\beta}+pqe^{i(\alpha+\beta)}\\
pe^{i\alpha}+qe^{i\beta}+pqe^{i(\alpha+\beta)} & pe^{i\alpha}+qe^{i\beta}+pqe^{i(\alpha+\beta)} &
 1+p^2e^{2i\alpha}+q^2e^{2i\beta}\\
           \end{array}\right)\label{efmnu}
\end{equation}
where we have parametrized the different elements of $m_\nu$ in terms of two real parameters $p$, $q$ 
and two phase parameters $\alpha$, $\beta$ accordingly as
\begin{eqnarray}
m_0=-\frac{y_3^2}{m} ,
\quad pe^{i\alpha}=\frac{y_1}{y_3},
\quad qe^{i\beta}=\frac{y_2}{y_3}.\label{parametrization}
\end{eqnarray}
Upon diagonalization, $m_\nu$
yields degenerate mass eigenvalues\cite{Adhikary:2013bma}. 
The eigenvectors corresponding to the degenerate eigenvalues can not be determined uniquely. Hence the Diagonalization
matrix is not unique and there by generates an ambiguity\footnote{We put a brief explanation of this ambiguity in Ref.\cite{Adhikary:2013bma}.} 
in the neutrino mixing angles.
Thus, it is necessary to break the discrete symmetry in order to accommodate neutrino oscillation data.
Retaining the flavor diagonal texture of $M_R$, we introduce single nonzero 
symmetry breaking parameters $\epsilon$ in any of the diagonal entries. 
This can be done in three ways as\\
\begin{eqnarray}
&&(i)~M_R=diag(m,~m,~m(1+\epsilon)) \nonumber\\
&&(ii)~M_R=diag(m,~m(1+\epsilon),~m)\nonumber\\&&(iii)~M_R=diag(m(1+\epsilon),~m,~m) ~~.\label{3cat_brk}
\end{eqnarray}
The $m_\nu$ matrices for the above mentioned three cases of symmetry breaking are (using the unique parametrization shown 
in eq.(\ref{parametrization})) given by\\\\
for case(i)
\begin{equation}
m_\nu=m_0 \left(
\begin{array}{ccc}
 e^{2 i \alpha } p^2+e^{2 i \beta } q^2+\frac{1}{1+\epsilon } & e^{i \alpha } p+e^{i (\alpha +\beta )} p q+\frac{e^{i \beta } q}{1+\epsilon } & e^{i \beta } q+e^{i (\alpha +\beta )} p q+\frac{e^{i \alpha } p}{1+\epsilon } \\
 e^{i \alpha } p+e^{i (\alpha +\beta )} p q+\frac{e^{i \beta } q}{1+\epsilon } & 1+e^{2 i \alpha } p^2+\frac{e^{2 i \beta } q^2}{1+\epsilon } & e^{i \alpha } p+e^{i \beta } q+\frac{e^{i (\alpha +\beta )} p q}{1+\epsilon } \\
 e^{i \beta } q+e^{i (\alpha +\beta )} p q+\frac{e^{i \alpha } p}{1+\epsilon } & e^{i \alpha } p+e^{i \beta } q+\frac{e^{i (\alpha +\beta )} p q}{1+\epsilon } & 1+e^{2 i \beta } q^2+\frac{e^{2 i \alpha } p^2}{1+\epsilon }
\end{array}
\right) \label{mnu1} 
\end{equation}\\
for case(ii)
\begin{equation}
 m_\nu=m_0 \left(
\begin{array}{ccc}
 1+e^{2 i \alpha } p^2+\frac{e^{2 i \beta } q^2}{1+\epsilon } & e^{i \alpha } p+e^{i \beta } q+\frac{e^{i (\alpha +\beta )} p q}{1+\epsilon } & e^{i \alpha } p+e^{i (\alpha +\beta )} p q+\frac{e^{i \beta } q}{1+\epsilon } \\
 e^{i \alpha } p+e^{i \beta } q+\frac{e^{i (\alpha +\beta )} p q}{1+\epsilon } & 1+e^{2 i \beta } q^2+\frac{e^{2 i \alpha } p^2}{1+\epsilon } & e^{i \beta } q+e^{i (\alpha +\beta )} p q+\frac{e^{i \alpha } p}{1+\epsilon } \\
 e^{i \alpha } p+e^{i (\alpha +\beta )} p q+\frac{e^{i \beta } q}{1+\epsilon } & e^{i \beta } q+e^{i (\alpha +\beta )} p q+\frac{e^{i \alpha } p}{1+\epsilon } & e^{2 i \alpha } p^2+e^{2 i \beta } q^2+\frac{1}{1+\epsilon }
\end{array}
\right) \label{mnu2}
\end{equation}\\
for case(iii)
\begin{equation}
m_\nu=m_0\left(
\begin{array}{ccc}
 1+e^{2 i \beta } q^2+\frac{e^{2 i \alpha } p^2}{1+\epsilon } & e^{i \beta } q+e^{i (\alpha +\beta )} p q+\frac{e^{i \alpha } p}{1+\epsilon } & e^{i \alpha } p+e^{i \beta } q+\frac{e^{i (\alpha +\beta )} p q}{1+\epsilon } \\
 e^{i \beta } q+e^{i (\alpha +\beta )} p q+\frac{e^{i \alpha } p}{1+\epsilon } & e^{2 i \alpha } p^2+e^{2 i \beta } q^2+\frac{1}{1+\epsilon } & e^{i \alpha } p+e^{i (\alpha +\beta )} p q+\frac{e^{i \beta } q}{1+\epsilon } \\
 e^{i \alpha } p+e^{i \beta } q+\frac{e^{i (\alpha +\beta )} p q}{1+\epsilon } & e^{i \alpha } p+e^{i (\alpha +\beta )} p q+\frac{e^{i \beta } q}{1+\epsilon } & 1+e^{2 i \alpha } p^2+\frac{e^{2 i \beta } q^2}{1+\epsilon }
\end{array}
\right) ~~.\label{mnu3}
\end{equation}

All the experimentally measurable observable (mass squared differences and mixing angles) of this broken symmetric 
neutrino mass matrix are obtained in terms of the Lagrangian parameters ($p$, $q$, $\alpha$, $\beta$, $m_0$) and 
breaking parameter($\epsilon$) using the methodology developed in Ref.\cite{Adhikary:2013bma} to calculate
the masses and mixing angles from the most general Majorana neutrino mass matrix.
\section{Baryogenesis through leptogenesis}\label{baryo}
Here we will discuss about the lepton asymmetry arising from a CP asymmetry\cite{Covi:1996wh} generated due to the decay of heavy
 right handed Majorana
neutrinos. At a high energy scale where $SU(2)_L\times U(1)_Y$ symmetry is not broken, physical
right handed neutrinos $N_{iR}$ with definite mass can decay into (i) charged lepton with charged scalar
and (ii) light neutrino with neutral scalar. The conjugate decay process is also possible due to
self conjugate nature of $N_{iR}$. A net lepton asymmetry will be generated if two decay processes 
occur at different rate. 
In the present case the right handed Majorana neutrinos are not hierarchical. 
Before the explicit breaking of the cyclic symmetry
the mass spectrum of the right handed neutrinos is degenerate. 
After cyclic symmetry breaking the masses of the heavy right 
handed neutrinos differ by the symmetry breaking parameter($\epsilon$).
~Primarily we have varied the symmetry breaking parameter ($\epsilon$) within a large range 
in order to fit neutrino oscillation data. With the motivation of keeping the symmetry breaking soft as well as to study
the leptogenesis behaviour for quasi degenerate right handed neutrinos we pick only the smallest value of $\epsilon$ ($\sim 0.004$)
~allowed by the oscillation data.
In this case the right handed neutrinos are nearly degenerate
and there is a possibility of occurrence of resonant leptogenesis\footnote{Only the smallness of
the breaking parameter doesn't guarantee the resonance enhancement of CP asymmetry, resonance occurs only
when the resonance condition is satisfied.} in this situation. This is the less addressed interesting
case of leptogenesis, which we study in the present work. For higher
values of $\epsilon$ ($\sim 0.1$) right handed neutrinos are hierarchical and leptogenesis phenomena
in this case has been well studied in the literature. In this work we explore the parameter region where
the leptogenesis takes place due to decay of quasi degenerate right handed neutrinos.
Unlike the strongly hierarchical case here we have to consider contributions from all 
three generations of right handed neutrinos \cite{Pilaftsis:2003gt} to calculate the CP asymmetry parameters. \\
\subsection{Calculation of CP asymmetry parameters}
The resummed effective Yukawa couplings (considering three
generations of $N_{iR}$) are given by\cite{Pilaftsis:2003gt,Pilaftsis:1997jf}
\begin{eqnarray}
  \label{hres3g}
(\bar{h}^\nu_+ )_{\alpha i} \!&=&\! 
h^\nu_{\alpha i}\, +\, iB_{\alpha i}\: -\: i\, \sum_{j,k=1}^3\,
|\epsilon_{ijk}|\, h^\nu_{\alpha j}\nonumber\\
&&\hspace{-2cm}\times\,\frac{m_{N_i} ( m_{N_i} A_{ij} + m_{N_j} A_{ji}) 
+ R_{ik} \Big[ m_{N_i} A_{kj} ( m_{N_i} A_{ik} + m_{N_k} A_{ki} )
+ m_{N_j} A_{jk} ( m_{N_i} A_{ki} + m_{N_k} A_{ik} ) \Big]}
{ m^2_{N_i}\, -\, 
m^2_{N_j}\, +\, 2i\,m^2_{N_i} A_{jj} + 2i\,{\rm Im}R_{ik}\,
\Big( m^2_{N_i} |A_{jk}|^2 + 
m_{N_j} m_{N_k} {\rm Re}A^2_{jk}\Big)   }\ ,\nonumber\\
\end{eqnarray}
where
\begin{eqnarray}
&&R_{ij}\ =\ \frac{m^2_{N_i}}{m^2_{N_i} - m^2_{N_j} + 2i\, m^2_{N_i} 
A_{jj}},\\
&&A_{ij}=\frac{({h^\nu}^\dagger h^\nu)_{ji}}{16\pi},\\
&&B_{\alpha i}=-\sum_{(j\neq i)}\frac{({h^\nu}^\dagger h^\nu)_{ij}h^\nu_{\alpha j}}{16\pi}f(\frac{m_{N_j}^2}{p^2}),
\end{eqnarray}
$(h^\nu)_{\alpha i}$ is tree level neutrino Yukawa coupling and $|\epsilon_{ijk}|$ is   the modulus of the usual   Levi--Civita
anti-symmetric tensor. \\\\
The resummed effective amplitudes for the decays $N_{iR} \rightarrow l_\alpha \Phi$ are denoted as 
${\cal T} (N_{iR} \to l_\alpha \Phi)$ and are given by 
\begin{equation}
  \label{Thres}
{\cal T} (N_{iR} \to l_\alpha\Phi) \ =\ 
 (\bar{h}^\nu_+)_{\alpha i}\ \bar{u}_\alpha\, P_R\, u_{N_i}\;, 
\end{equation}
where $i$ ($i=1,2,3$) and $\alpha$ ($\alpha=e,\mu,\tau$) are the generation indices of $N_{iR}$ and leptons respectively
and ${u}_\alpha$, $u_{N_i}$ denote corresponding spinorial fields. The CP conjugate decay amplitudes 
${\cal T} (N_{iR} \to l_\alpha^c \Phi^\dagger)$ can be obtained easily from eq.(\ref{Thres}) by replacing
 $(\bar{h}^\nu_+)_{\alpha i}$ with
$(\bar{h}^\nu_-)_{\alpha i}$ which can be further recovered from eq.(\ref{hres3g}) by taking complex conjugate 
of the Yukawa couplings.
The CP asymmetry of the decay is characterized by a parameter $\varepsilon^\alpha_i$ defined by
\begin{eqnarray}
\varepsilon^\alpha_i &=&\frac{\Gamma(N_{iR}\rightarrow l_\alpha \Phi)-
\Gamma(N_{iR}\rightarrow l_\alpha^c \Phi^\dagger)}
{\Sigma_\alpha[\Gamma(N_{iR}\rightarrow l_\alpha \Phi)+\Gamma(N_{iR}\rightarrow l_\alpha^c \Phi^\dagger)]}\nonumber\\
&=&\frac{(\bar{h}^{\nu\,\dagger}_+)_{i\alpha }( \bar{h}^\nu_+)_{\alpha i} \: - \: 
(\bar{h}^{\nu\,\dagger}_-)_{i\alpha }( \bar{h}^\nu_-)_{\alpha i}}{
(\bar{h}^{\nu\,\dagger}_+ \bar{h}^\nu_+)_{ii} \: + \: 
(\bar{h}^{\nu\,\dagger}_- \bar{h}^\nu_-)_{ii}}\ \label{cpasy} .
\end{eqnarray}
After a long algebraic manipulation the 
expression of $\varepsilon^\alpha_i$ is presented in a simpler form keeping terms upto $O({h^\nu}^4)$ as
\begin{eqnarray}
\varepsilon^\alpha_i
&=&\frac{1}{4\pi v^2 H_{ii}}\sum_{j\ne i} Im\{H_{ij}(m_D^\dagger)_{i\alpha} (m_D)_{\alpha j}\}
\left[f(x_{ij})+\frac{\sqrt{x_{ij}}(1-x_{ij})}
{(1-x_{ij})^2+\frac{H_{jj}^2}{16 \pi^2 v^4}}\right]\nonumber\\
&+&\frac{1}{4\pi v^2 H_{ii}}\sum_{j\ne i}\frac{(1-x_{ij})Im\{H_{ji}(m_D^\dagger)_{i\alpha} (m_D)_{\alpha j}\}}
{(1-x_{ij})^2+\frac{H_{jj}^2}{16 \pi^2 v^4}}
\label{epsi}
\end{eqnarray}
where  $m_D=\frac{vh^\nu}{\sqrt{2}}$, $H=(m_D^\dagger m_D)$, $x_{ij}=\frac{m_{N_j}^2}{m_{N_{i}}^2}$ and 
 $f(x_{ij})$ is the loop function given by
\begin{equation}
f(x_{ij})=\sqrt{x_{ij}}\{1-(1+x_{ij})\ln(\frac{1+x_{ij}}{x_{ij}})\}.
\end{equation}
Again, derived expressions for $\varepsilon^\alpha_{i}$ 
are quite general and can be used for  hierarchical as well as  quasi degenerate mass spectrum 
(without or with resonant conditions like $1-x_{ij}\simeq \frac{H_{jj}}{4\pi v^2}$ ) of right handed neutrinos.
For hierarchical case one can simplify $\varepsilon^\alpha_{i}$ to standard formula\cite{Adhikary:2010fa}
neglecting $\frac{H_{jj}^2}{16\pi^2 v^4}$ compared to $(1-x_{ij})^2$. We have neglected  $O({h^\nu}^6)$ and higher 
order terms in our obtained expression of $\varepsilon^\alpha_i$ in eq.(\ref{epsi}). Contribution of those terms are negligible
for most of the cases. 
\subsection{Boltzmann equations for leptogenesis}
In the present work the right handed neutrinos are taken to be nearly degenerate. Thus in a temperature regime
where the lepton flavors are distinguishable, we have to consider the flavor dependent as well as resonant leptogenesis
formalism. The corresponding set of Boltzmann equations are given by\cite{Pilaftsis:2003gt}
\begin{eqnarray}
  \label{BEN} 
\frac{d \eta_{N_i}}{dz} &=& \frac{z}{H(z=1)}\ \bigg[\,\bigg( 1
\: -\: \frac{\eta_{N_i}}{\eta^{\rm eq}_{N_i}}\,\bigg)\,\sum\limits_{\alpha} \bigg(\,
\Gamma^{\alpha ~D\; (i)} \: +\: \Gamma^{\alpha ~S\; (i)}_{\rm Yukawa}\: +\:
\Gamma^{\alpha ~S\; (i)}_{\rm Gauge}\, \bigg) \nonumber\\ 
&&-\, \frac{1}{4}\,\sum\limits_{\alpha} \eta^\alpha_L\, \varepsilon^\alpha_i\, \bigg(\, \Gamma^{\alpha ~D\; (i)} \: +\:
\widetilde{\Gamma}^{\alpha ~S\; (i)}_{\rm Yukawa}\: +\:
\widetilde{\Gamma}^{\alpha ~S\; (i)}_{\rm Gauge}\, \bigg)\,\bigg]\,,\\[3mm] 
  \label{BEL}
\frac{d \eta^\alpha_L}{dz} &=& -\, \frac{z}{H(z=1)}\, \bigg\{\,
\sum\limits_{i=1}^3\,\varepsilon^\alpha_i \ 
\bigg( 1 \: -\: \frac{\eta_{N_i}}{\eta^{\rm eq}_{N_i}}\,\bigg)\,\sum\limits_{\beta} \bigg(\,
\Gamma^{\beta ~D\; (i)} \: +\: \Gamma^{\beta ~S\; (i)}_{\rm Yukawa}\: +\:
\Gamma^{\beta ~S\; (i)}_{\rm Gauge}\, \bigg) \nonumber\\ 
&&+\, \frac{1}{4}\, \eta^\alpha_L\, \bigg[\, \sum\limits_{i=1}^3\, 
\bigg(\, \Gamma^{\alpha ~D\; (i)} \: +\: 
\Gamma^{\alpha ~W\;(i)}_{\rm Yukawa}\: 
+\: \Gamma^{\alpha ~W\; (i)}_{\rm Gauge}\,\bigg)\: +\:
\Gamma^{\alpha ~\Delta L =2}_{\rm Yukawa} \bigg]\,\bigg\}\,,
\end{eqnarray}
where\\ $z=\frac{mass ~of~ lightest~ right ~handed~ neutrino}{temperature}=\frac{m_{N_1}}{T}$ and the parameter $\eta_a$
give the number density of a particle species $a$ normalized to the photon density, i.e $\eta_a(z)=\frac{n_a(z)}{n_\gamma(z)}$
and $\eta^{\rm eq}_a(z)=\frac{n^{\rm eq}_a(z)}{n_\gamma(z)}$
with $n_\gamma(z)=\frac{2 m_{N_1}^3}{\pi^2 z^3}$. 
The number density of a particle species $a$ with  $g_a$ internal
degrees of freedom is given by
\begin{eqnarray}
  \label{na}
n_a (T)=  \frac{g_a\, m^2_a\,T\ e^{\mu_a (T)/T}}{2\pi^2}\
K_2\bigg(\frac{m_a}{T}\bigg)\; 
\end{eqnarray}
and $n_a (T)$ satisfies the equilibrium density condition when $\mu_a  = 0$, i.e\cite{MAL}
\begin{equation}
n^{\rm eq}_a (T)=  \frac{g_a\, m^2_a\,T\ }{2\pi^2}\
K_2\bigg(\frac{m_a}{T}\bigg)\; .
\end{equation}
Here $a$ denotes a definite particle species.
The various $\Gamma$ s in the RHS of the Boltzmann equations are normalized (by photon density) decays and scattering 
cross sections\cite{Pilaftsis:2003gt}, 
\begin{eqnarray}
  \label{GD}
&&\Gamma^{\alpha~ D\; (i)}  =  \frac{1}{n_\gamma}\ \gamma^{N_i}_{L^\alpha \Phi}\;, \nonumber\\
  \label{GSY}
&&\Gamma^{\alpha~ S\; (i)}_{\rm Yukawa}  = \frac{1}{n_\gamma}\
\bigg(\, \gamma^{N_i L^\alpha}_{Q u^C}\: +\:  \gamma^{N_i u^C}_{L^\alpha Q^C}\: 
+\: \gamma^{N_i Q}_{L^\alpha u}\, \bigg)\; ,\nonumber\\
&&\widetilde{\Gamma}^{\alpha ~S\;(i)}_{\rm Yukawa} = \frac{1}{n_\gamma}\
\bigg(\, \frac{\eta_{N_i}}{\eta^{\rm eq}_{N_i}}\, \gamma^{N_i L^\alpha}_{Q u^C}\: 
+\: \gamma^{N_i u^C}_{L^\alpha Q^C}\: +\: \gamma^{N_i Q}_{L^\alpha u}\, \bigg)\;,\nonumber\\
  \label{GSG}
&&\Gamma^{\alpha ~S\; (i)}_{\rm Gauge}  =  \frac{1}{n_\gamma}\ 
\bigg(\, \gamma^{N_i V_\mu}_{L^\alpha ~\Phi}\: +\: 
\gamma^{N_i L^\alpha}_{\Phi^\dagger V_\mu}\: +\: 
\gamma^{N_i\Phi^\dagger }_{L^\alpha V_\mu}\, \bigg)\;,\nonumber\\
&&\widetilde{\Gamma}^{\alpha ~S\; (i)}_{\rm Gauge} = \frac{1}{n_\gamma}\ 
\bigg(\, \gamma^{N_i V_\mu}_{L^\alpha \Phi}\: +\: 
\frac{\eta_{N_i}}{\eta^{\rm eq}_{N_i}}\, 
\gamma^{N_i L^\alpha}_{\Phi^\dagger V_\mu}\: +\: 
\gamma^{N_i\Phi^\dagger }_{L^\alpha V_\mu}\, \bigg)\; ,\nonumber\\
  \label{GWY}
&&\Gamma^{\alpha ~W\; (i)}_{\rm Yukawa}  =  \frac{2}{n_\gamma}\
\bigg(\, \gamma^{N_i L^\alpha}_{Q u^C}\: +\:  \gamma^{N_i u^C}_{L^\alpha Q^C}\: 
+\: \gamma^{N_i Q}_{L^\alpha u}\: +\: \frac{\eta_{N_i}}{2\eta^{\rm eq}_{N_i}}\,
\gamma^{N_i L^\alpha}_{Q u^C}\, \bigg)\; ,\nonumber\\
  \label{GWG}
&&\Gamma^{\alpha ~W\; (i)}_{\rm Gauge}  =  \frac{2}{n_\gamma}\ 
\bigg(\, \gamma^{N_i V_\mu}_{L^\alpha \Phi}\: +\: 
\gamma^{N_i L^\alpha}_{\Phi^\dagger V_\mu}\: +\: 
\gamma^{N_i\Phi^\dagger }_{L^\alpha V_\mu}\: +\:
\frac{\eta_{N_i}}{2\eta^{\rm eq}_{N_i}}\, 
\gamma^{N_i L^\alpha}_{\Phi^\dagger V_\mu}\, \bigg)\;,\nonumber\\
  \label{GDL2}
&&\Gamma^{\alpha ~\Delta L =2}_{\rm Yukawa} = \frac{2}{n_\gamma}\sum_\beta\ 
\bigg(\, \gamma^{\,\prime L^\alpha\Phi}_{\,L^{\beta C}\Phi^\dagger} 
+\:  2\gamma^{L^\alpha L^\beta}_{\Phi^\dagger\Phi^\dagger}\, \bigg)\; 
\end{eqnarray}
where $\alpha$ denotes lepton flavor indices ($e,~\mu,~\tau$).
For a generic process $X~\rightarrow~Y$, $\gamma^X_Y$ is defined as
\begin{equation}
  \label{CT}
\gamma^X_Y\ \equiv \ \gamma ( X\to Y)\: +\: \gamma ( \overline{X}
\to \overline{Y} )\; ,
\end{equation}
with
\begin{equation}
  \label{gamma}
\gamma ( X\to Y)\ =\ \int\! d\pi_X\, d\pi_Y\, (2\pi )^4\,
\delta^{(4)} ( p_X - p_Y )\ e^{-p^0_X/T}\, |{\cal M}( X \to Y )|^2\; .
\end{equation}
The explicit expressions of different $\gamma$ s listed above are taken from the appendix of \cite{Pilaftsis:2003gt}
\footnote{Those expressions are free of flavor index $\alpha$. We have introduced the flavor indices in suitable places.}.
\\\\
When the resonance condition ($1-x_{ij} =\frac{H_{jj}}{4\pi v^2}$) is not satisfied strictly, such that the 
enhancement of the CP asymmetry parameter $(\varepsilon^\alpha_i)$ is not too high, contribution of the 
second term of eq.(\ref{BEN}) is negligible compared to the first term. In that case the first Boltzmann equation (\ref{BEN})
can be rewritten in a simpler form as
\begin{equation}
\frac{d \eta_{N_i(z)}}{d z}=(D_i(z)+D^{\rm SY}_i(z)+D^{\rm SG}_i(z))(\eta^{\rm eq}_{N_i}(z)-\eta_{N_i}(z)).\label{BEN1}
\end{equation}
In terms of another parameter $Y_{N_i}$ ($=n_{N_i}/s$, where $n_{N_i}$ is the number density of $N_i$ and $s$ is the 
comoving entropy density )
\footnote{$\eta_{N_i(z)}=1.8g_{\ast s}(T)Y_{N_i}(z)$, but in our regime of interest $g_{\ast s}(T)$
is nearly constant and thus we can say that $\eta_{N_i(z)}$ and $Y_{N_i}(z)$ are connected through a constant factor.}
the above equation can be rewritten as
\begin{equation}
\frac{d Y_{N_i(z)}}{d z}=(D_i(z)+D^{\rm SY}_i(z)+D^{\rm SG}_i(z))(Y^{\rm eq}_{N_i}(z)-Y_{N_i}(z))\label{BEN_Y1}
\end{equation}
where
\begin{eqnarray}
&&D_i(z)=\sum \limits_\alpha D^\alpha_i(z) = \sum \limits_\alpha \frac{z}{H(z=1)}\frac{\Gamma^{\alpha~ D\; (i)}}{\eta^{\rm eq}_{N_i}(z)},\nonumber\\
&&D^{\rm SY}_i(z)=\sum \limits_\alpha \frac{z}{H(z=1)}\frac{\Gamma^{\alpha~ S\; (i)}_{\rm Yukawa}}{\eta^{\rm eq}_{N_i}(z)},\nonumber\\
&&D^{\rm SG}_i(z)=\sum \limits_\alpha \frac{z}{H(z=1)}\frac{\Gamma^{\alpha~ S\; (i)}_{\rm Gauge}}{\eta^{\rm eq}_{N_i}(z)}\nonumber.
\end{eqnarray}
Similarly (neglecting $\Delta L=2$ scattering terms) the second Boltzmann equation (\ref{BEL}) can be written as
\begin{eqnarray}
\frac{d \eta^\alpha_L(z)}{dz} = & -&\{ \sum\limits_{i=1}^3\,\varepsilon^\alpha_i \ (D_i(z)+D^{\rm SY}_i(z)+D^{\rm SG}_i(z))(\eta^{\rm eq}_{N_i}(z)-\eta_{N_i}(z))
\nonumber\\& + & \frac{1}{4}\eta^\alpha_L\sum\limits_{i=1}^3\ 
(\frac{1}{2}D^\alpha_i(z)z^2 K_2(z)+D^{\alpha ~\rm YW}_i(z)+ D^{\alpha ~\rm GW}_i(z) ) \} \label{BEL1}
\end{eqnarray}
with
\begin{eqnarray}
&&D^{\rm YW}_i(z)=\sum \limits_\alpha \frac{z}{H(z=1)}\Gamma^{\alpha~ W\; (i)}_{\rm Yukawa},\nonumber\\
&&D^{\rm GW}_i(z)=\sum \limits_\alpha \frac{z}{H(z=1)}\Gamma^{\alpha~ W\; (i)}_{\rm Gauge}\nonumber.
\end{eqnarray}
\\
The second Boltzmann equation governs the evolution of the lepton flavor asymmetry ($\eta^\alpha_L$). Now we will
discuss the recipe to calculate the baryon asymmetry from lepton flavor asymmetry. At first we introduce a parameter
$Y_\alpha$ which is number density (of a lepton flavor) normalized by entropy density ($s$) and it is related to $\eta^\alpha_L$
through
\begin{equation}
Y_\alpha=\frac{n^\alpha_L -n^\alpha_{\overline{L}}}{s}=(\frac{\eta_\gamma}{s})\eta^\alpha_L.    
\end{equation}
We know that $\frac{s}{\eta_\gamma}=1.8g_{\ast s}$\cite{kolb_turner} where $g_{\ast s}$ counts total number of effective massless degrees
of freedom and it is a function of temperature. For $T>10^2~{\rm GeV}$, $g_{\ast s}$ is nearly constant and its value 
(with three right handed neutrinos) is $112$\cite{Adhikary:2006rf}.
The lepton asymmetry created by the decay of right handed neutrinos is converted into baryon asymmetry through sphaleron process.
During the sphaleron process the quantity $\Delta_\alpha=\frac{B}{3}-L^\alpha$ (where $B$ is the baryon number 
and $L$ is the lepton number) is conserved. The $Y_{\Delta_\alpha}$
asymmetries and $Y_\alpha$ asymmetries are related through a matrix \textquoteleft$A$\textquoteright as 
$Y_\alpha=\sum\limits_{\beta}A_{\alpha\beta}Y_{\Delta_\beta}$.
The Boltzmann equation (\ref{BEL1}) governing the evolution of flavor asymmetries can be written in terms of $Y_{\Delta_\alpha}$
parameters as
\begin{eqnarray}
\frac{d Y_{\Delta_\alpha}}{dz} = & \sum\limits_{i=1}^3 \, & \{\varepsilon^\alpha_i \ (D_i(z)+D^{\rm SY}_i(z)+D^{\rm SG}_i(z))(Y^{\rm eq}_{N_i}(z)-Y_{N_i}(z))
\}  \nonumber\\ & + &\frac{\sum\limits_{\beta}A_{\alpha\beta}Y_{\Delta_\beta}}{4} \{\sum\limits_{i=1}^3\ 
(\frac{1}{2}D^\alpha_i(z)z^2 K_2(z)+D^{\alpha ~\rm YW}_i(z)+ D^{\alpha ~\rm GW}_i(z) )\}  .\label{BEL_Y1}
\end{eqnarray}
We now solve the set of coupled Boltzmann equations (given in eq.(\ref{BEN_Y1}) and eq.(\ref{BEL_Y1})) upto a value of $z$ where the quantities
$Y_{\Delta_\alpha}$ attain a constant value.
\section{Calculation of baryon asymmetry in different energy regimes}\label{baryo1}
We are now in a position 
to compute the baryon asymmetry in different regimes\cite{k3,antush}.
\subsubsection{${\bf{M_{lowest}<{10}^{9}}}$ GeV.}\label{ff}
Here all three lepton flavors are separately active. Thus the $A$ matrix connecting $Y_{\Delta_\alpha}$
and $Y_\alpha$ is a $3 \times 3$ matrix given by\cite{k3}
\begin{equation}
A=\left(\begin{array}{ccc}
-151/179 & 20/179 & 20/179\\
25/358 & -344/537 & 14/537\\
25/358 &  14/537 &  -344/537
\end{array}\right) .
\end{equation}
The final baryon asymmetry $Y_B$ (baryon asymmetry normalized by entropy density) is given by\cite{harvey}
\begin{equation}
Y_B=\frac{28}{79}\sum\limits_{\alpha} Y_{\Delta_\alpha} .
\end{equation}
Another important parameter, i.e baryon asymmetry normalized to photon
density is obtained as
\begin{equation}
\eta_B=\left.\frac{s}{n_\gamma}\right|_0Y_B=7.0394Y_B,
\end{equation}
where $\eta_B=\left.\frac{s}{n_\gamma}\right|_0Y_B=7.0394Y_B$, the zero at the subscript denotes its value at present epoch.

\subsubsection{ ${\bf{{10}^{9}\,\,{\rm{\bf{{\rm GeV}}}}<M_{lowest}<{10}^{12}}}$ GeV.}\label{pf}
In this regime $\tau$ flavor is distinguishable but we can't 
differentiate between $e$ and $\mu$ flavors. So we define two CP asymmetries $\varepsilon^\tau_{N_i}$ and 
$\varepsilon^2_{N_i}=\varepsilon^e_{N_i}+\varepsilon^\mu_{N_i}$ and the corresponding lepton flavor asymmetry
parameters are $Y_\tau$ and $Y_2=Y_e+Y_\mu$. Solving the Boltzmann equations (\ref{BEN_Y1},\ref{BEL_Y1})
we get two $Y_\Delta$ asymmetries ($Y_{\Delta_2}$ and $Y_{\Delta_\tau}$) and the final baryon asymmetry parameter is calculated as
\begin{equation}
Y_B=\frac{28}{79}(Y_{\Delta_2}+Y_{\Delta_\tau}).
\end{equation}
In this case the $A$ matrix connecting $Y_\alpha$ and $Y_{\Delta_\alpha}$ is a $2 \times 2$ matrix given by\cite{k3}
\begin{equation}
A=\left(\begin{array}{cc}
  -417/589  & 120/589\\
   30/589 & -390/589
\end{array}\right). 
\end{equation}

\subsubsection{ ${\bf{M_{lowest}>{10}^{12}}}$ GeV.}\label{uf}
All the charged lepton flavors act indistinguishably in this regime and 
therefore one can define a single CP asymmetry parameter $\varepsilon_i=\sum\limits_{\alpha}\varepsilon^\alpha_i$.
The other $\alpha$ dependent terms in RHS of eq.(\ref{BEL_Y1}) are replaced by sum over $\alpha$ and the 
$A$ matrix is taken as negative unit matrix.
So the Boltzmann equation (\ref{BEL_Y1}) is now free of the flavor index $\alpha$ and solving
the same we get a single $Y_\Delta$. The final baryon asymmetry parameter is obtained as
\begin{equation}
Y_B= \frac{28}{79} Y_\Delta.
\end{equation}

\section{Numerical results and phenomenological discussion}\label{nu_ph}
For numerical estimation of  baryon asymmetry we need to know 
CP asymmetry parameters $\varepsilon^\alpha_{i}$ and various decays and scattering cross sections in terms of the parameters
$m,~m_0,~p,~q,~\alpha,~\beta$ and $\epsilon$.
However, dependencies of those parameters on  $\varepsilon^\alpha_{i}$ and the decay/scattering terms arise through the expressions of
$m_D$, $H(=m_D^\dagger m_D)$ and $x_{ij}$. Obviously it is then necessary to express, $m_D$, $H$ and $x_{ij}$ in terms of those 
Lagrangian parameters. 
Utilizing eq.(\ref{parametrization}) we explicitly express the elements of $m_D$ in terms of the aforesaid parameters through
\begin{eqnarray}
\label{param1}
 y_3=i\sqrt{mm_0},\quad y_1 =i\sqrt{mm_0}pe^{i\alpha},\quad{\rm and }\quad y_2=i\sqrt{mm_0}qe^{i\beta}
\end{eqnarray}
which in effect gives
\begin{eqnarray}
\label{parammdH}
m_D&=&i\sqrt{mm_0}\left( \begin{array}{ccc}
      pe^{i\alpha} & qe^{i\beta} & 1\\1 & pe^{i\beta} & qe^{i\beta} \\ qe^{i\beta} & 1 & pe^{i\alpha} \\
    \end{array}\right), \nonumber\\
H&=&m_D^\dagger m_D\nonumber\\
&=& \left( \begin{array}{ccc}
     |y_1|^2+|y_2|^2+|y_3|^2 & y_1^*y_2+y_1y_3^*+y_2^*y_3 &y_1y_2^*+y_1^*y_3+y_2y_3^*\\ y_1y_2^*+y_1^*y_3+y_2y_3^* &
 |y_1|^2+|y_2|^2+|y_3|^2 & y_1^*y_2+y_1y_3^*+y_2^*y_3 
\\ y_1^*y_2+y_1y_3^*+y_2^*y_3 & y_1y_2^*+y_1^*y_3+y_2y_3^*& |y_1|^2+|y_2|^2+|y_3|^2 \\
    \end{array}\right) \nonumber\\
 &=& mm_0\left( \begin{array}{ccc} X&Y&Y^*\\Y^*&X&Y\\Y&Y^*&X\end{array}\right)\label{capH}
\end{eqnarray}
with
\begin{eqnarray}
\label{XY}
X&=&1+p^2+q^2\nonumber\\
Y&=&pe^{i\alpha}+qe^{-i\beta}+pqe^{i(\beta-\alpha)}.
\end{eqnarray}
Again $x_{ij}=m_{N_j}^2/m_{N_i}^2$ (and $x_{ji}=1/x_{ij}$) is estimated from $M_R$ as
\begin{eqnarray}
\label{xij}
&&x_{12}=1,~x_{23}=x_{13}=(1+\epsilon)^2~~{\rm for ~case(i)}\nonumber\\
&&x_{12}=\frac{1}{x_{23}}=(1+\epsilon)^2,~x_{13}=1~~{\rm for ~case(ii)}\nonumber\\
&&x_{12}=x_{13}=\frac{1}{(1+\epsilon)^2},~x_{23}=1~~{\rm for ~case(iii)}.
\end{eqnarray}
To find out the allowed parameter space we adopt the following methodology.
In the present work the parameter space is constrained due to the bound on the frozen value Baryon asymmetry 
parameter($\eta_B$ or $Y_B$)
keeping in mind all the neutrino oscillation experimental data(Table \ref{osc}). 
The parameter space is constrained in two
steps. At first all the neutrino physics observables (mass eigenvalues, mixing angles)
 are expressed in terms
of the Lagrangian parameters ($m_0$, $p$, $q$, $\alpha$, $\beta$) and the breaking parameter 
($\epsilon$)\footnote{Our parameter space satisfy i) the cosmological bound on sum mass 
$\sum m_i(=m_1+m_2+m_3)$ $<$ $(0.23 -1.11)eV$ \cite{Giusarma:2013pmn} with PLANCK \cite{Ade:2013zuv} and other cosmological
observations \cite{Bennett:2012zja}, \cite{Aihara:2011sj} ii) also the bound $|m_{\nu_{ee}}|$ $<$ $(0.14-0.38)eV$ 
\cite{Auger:2012ar} of neutrinoless double beta decay experiments \cite{Tortola:2012te,Giuliani:2010zz,Rodejohann:2012xd}. }.
In the first step the parameters get restricted by the experimental ranges of neutrino 
mass squared differences (solar and atmospheric) and mixing angles.
These constrained set of parameters are used thereafter to calculate the 
CP asymmetry parameters and hence the baryon asymmetry parameter $Y_B$ (or $\eta_B$) for 
different values of right handed neutrino mass $m$ (to take into account fully flavored, $\tau$-flavored and unflavored
leptogenesis).
Hence, the parameters get second round of
restriction from the limits on baryon asymmetry. 
\begin{table}[!ht]
\caption{Input data from neutrino oscillation experiments \label{osc} \cite{Tortola:2012te}}
\label{input}
\begin{center}
\begin{tabular}{|c|c|}
\hline
{ Quantity} & { $3\sigma$ ranges/other constraint}\\
\hline
$\Delta m_{21}^2$ & $7.12<\Delta m_{21}^2(10^{5}~ eV^{-2})<8.20$\\
$|\Delta m_{31}^2|(N)$ & $2.31<\Delta m_{31}^2(10^{3}~ eV^{-2})<2.74$\\
$|\Delta m_{31}^2|(I)$ & $2.21<\Delta m_{31}^2(10^{3}~ eV^{-2})<2.64$\\
$\theta_{12}$ & $31.30^\circ<\theta_{12}<37.46^\circ$\\
$\theta_{23}$ & $36.86^\circ<\theta_{23}<55.55^\circ$\\
$\theta_{13}$ & $7.49^\circ<\theta_{13}<10.46^\circ$\\
\hline
\end{tabular}
\end{center}
\end{table}
We have observed that apart from the two parameter breaking of cyclic symmetry($\epsilon_1$, $\epsilon_2 \neq 0$)
(which is studied extensively in Ref\cite{Adhikary:2013bma})
one parameter breaking  is also well fitted by the extant data.
In this work our main motivation is to study the resonance enhancement of the CP asymmetries. So we have tried to keep 
the masses of the right handed neutrinos as close as allowed by the oscillation data.
In order to pin down the parameter space for each type of leptogenesis we consider 
three categories of single parameter cyclic symmetry breaking designated by case(i), (ii) and (iii) in eq.(\ref{3cat_brk}).
\paragraph{}
Before going into the case wise details it is worthwhile to mention that 
\begin{enumerate}
\item{
We have studied the variation of CP asymmetry
parameters ($\varepsilon^\alpha_i$) with right handed neutrino mass ($m$) in all the cases of symmetry breaking mentioned
above. It is found that $|\varepsilon^\alpha_i|$ vs $m$ curve shows a resonance peak near $m=10^{12}$ GeV. Resonance
is achieved when the condition $(1-x_{ij})=\frac{H_{jj}}{4\pi v^2}$ is satisfied which gives
$m_{res}\simeq \frac{8\pi v^2\epsilon}{m_0}$ at the point of resonance.}
\item{
When the parameter space is constrained with the neutrino oscillation data it is seen that the value of $m_0$ decreases as the 
value (no matter positive/negative) of breaking parameter ($\epsilon$) is increased. 
Now the condition for resonance enhancement of CP asymmetry is given by $m_{res}\simeq \frac{8\pi v^2\epsilon}{m_0}$.
Therefore for a larger value of $\epsilon$ the mass of right handed neutrino required for resonance will also be bigger.
The value of $m_{res}$ is $\sim ~ 10^{12}$ GeV for the lowest allowed value of $\epsilon(=-0.004)$
\footnote{In the cases we have analyzed in section \ref{nu_ph} the resonance enhancement of CP asymmetry parameter 
takes place near $m \sim 10^{12}$ GeV, but we cannot see its effect in producing baryon asymmetry
since it is in the unflavored regime.}. So 
for any value of $|\epsilon|>0.004$, $m_{res}$ will also be greater than $10^{12}$ GeV which falls in the 
unflavored regime of leptogenesis and our breaking scheme and the imposed symmetry 
is such that it will produce a null asymmetry in this regime. Therefore even if we take a bigger value of $\epsilon$ we can't 
observe the effect of resonance since the corresponding $m_{res}$ is in the unflavored regime.
To get an $m_{res}$ in the flavored leptogenesis regime we have to take $|\epsilon|$ smaller than $0.004$ which is
again not allowed by the oscillation data. 
So we prefer to carry out the analysis in a region where the symmetry is softly broken.\\\\
It is worthwhile to make a small remark in this context. The maximum value of breaking parameter allowed
by the oscillation data (for case(i) of symmetry breaking) is $\epsilon=-0.78$ and for it $m_0 \sim 3.5\times10^{-4}$ eV. 
The right handed neutrino mass required for resonance comes out to be $m_{res} \sim 3.38\times10^{18}$ GeV
which is beyond the scope of fully flavored and $\tau$-flavored leptogenesis.  
}
\item
{To get an upper bound on right handed neutrino mass we use the perturbative unitarity limit\cite{Ghosal:2001ep}, i.e
\begin{eqnarray}
\frac{(h^\nu_{ij})^2}{4 \pi} & < & 1 \nonumber\\
\Rightarrow (m_D)^2_{ij} & < & 2\pi v^2.
\end{eqnarray}
According to our parametrization 
\begin{equation}
m=-\frac{y_3^2}{m_0}.
\end{equation}
Since $y_3$ is an element of $m_D$ matrix, we can write
\begin{equation}
|m|<\frac{2\pi v^2}{m_0}
\label{per_bou} 
\end{equation}
In our cases of interest this condition sets the upper limit of right handed neutrino mass near $10^{14}$ GeV.} 

\item{To solve Boltzmann equations we have considered two initial conditions $Y_{N_i}=0$ and $Y_{\Delta_\alpha}=0$. These
 mean that initially there were no lepton asymmetry and right handed neutrinos. Leptons at first produce  
appreciable amount of right handed  neutrinos which decay asymmetrically to leptons. Other conditions, frequently used in the literature,
are $Y_{N_i}=Y_{N_i}^{eq}$ and $Y_{\Delta_\alpha}=0$. But we restrict ourselves to the previous one to 
solve the equations.}
\end{enumerate}
\subsection{Numerical analysis for case(i) of symmetry breaking }
It is implemented by incorporating the symmetry breaking parameter $\epsilon$ in the \textquoteleft$33$\textquoteright~
element of $M_R$ as shown in case(i) of eq.(\ref{3cat_brk}).
The calculation of mixing angles and mass eigenvalues using the resulting
mass matrix (\ref{mnu1}) is carried out thereafter. The parameter space, constrained by the extant 
data, is used to find out the numerical value of the baryon asymmetry $Y_B$.
~By varying the mass of the right chiral neutrinos, we have studied leptogenesis in all three energy regimes
as mentioned earlier. 
\\\\
As mentioned earlier, we are interested in the study of enhancement of the CP asymmetry parameter for the
nearly degenerate right handed neutrinos we pick only those set of Lagrangian parameters $\{p$, $q$, $\alpha$, $\beta$, $m_0\}$ corresponding
to the lowest allowed (by oscillation data) modulus value of the breaking parameter $\epsilon$.
We proceed further to calculate the $Y_B$ for those restricted sets of values only.
After the first round of restriction (by oscillation data) it is found that the lowest allowed value
of the breaking parameter is $\epsilon=-0.004$ and for this value of $\epsilon$ we get $10$ sets of values
of the parameters $\{p,~q,~\alpha,~\beta,~m_0\}$. 
(There are two sets of values of $\{ p,~q,~\alpha,~\beta \}$ and for each set, there are five different values of 
$m_0$. Therefore all total we have ten sets of values which are shown in Table \ref{case1_osc}.) For this case (i)
normal hierarchy is preferred and $\theta_{23}$ is selected in the first octant ($37.64^\circ$). But the sign of 
$\alpha$, $\beta$  remain unsettled. This in effect produces sign ambiguity in Dirac CP phase: 
 $\delta_{\rm CP}=29.38^\circ$ for the set 1 with ($\alpha=-113.5^\circ$, $\beta=120.5^\circ$)  and
  $\delta_{\rm CP}=-29.38^\circ$ for the set 2 with ($\alpha=113.5^\circ$, $\beta=-120.5^\circ$). 

\begin{center}
\begin{figure}[!h]
\includegraphics[width=7.5cm,height=6.5cm,angle=0]{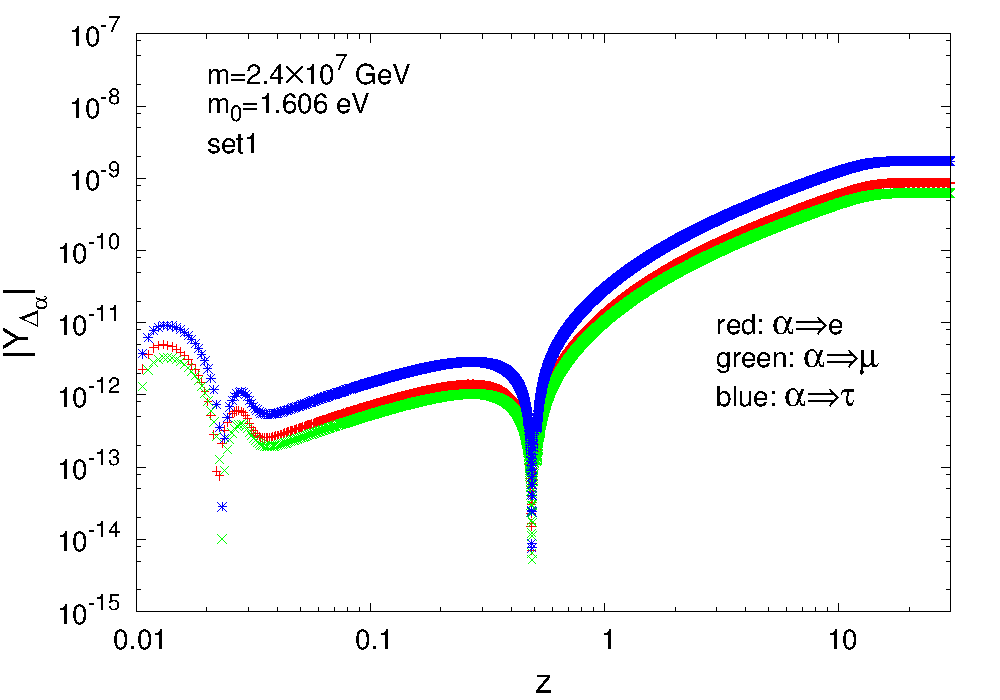}
\hspace{1.0cm}
\includegraphics[width=7.5cm,height=6.5cm,angle=0]{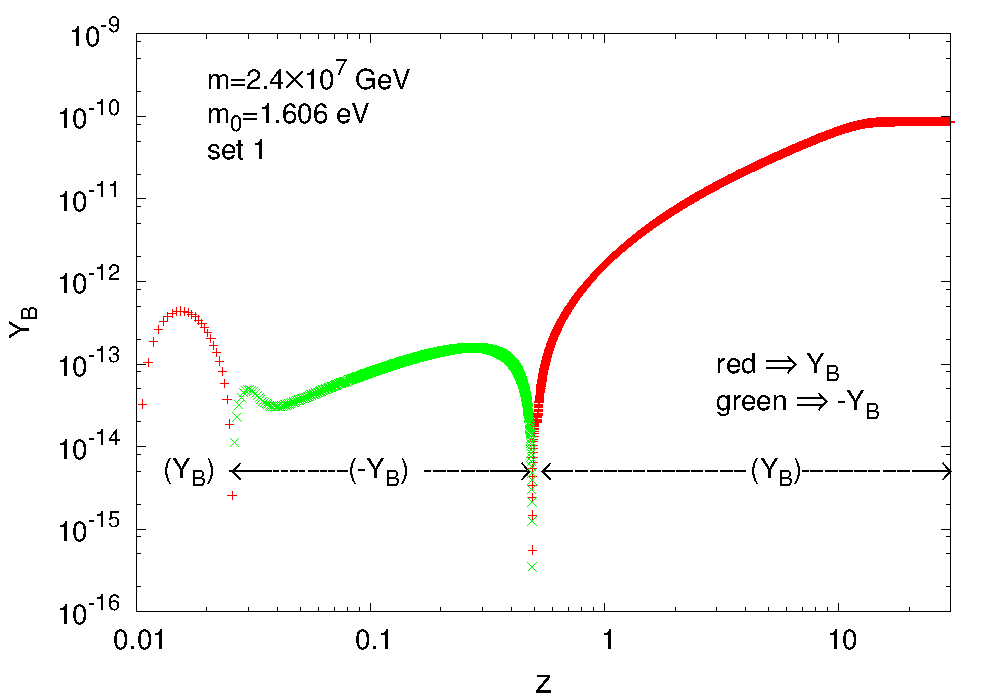}
\caption{(colour online) Plot of flavor asymmetries($Y_{\Delta_\alpha}$) (left) and baryon asymmetry($Y_B$) (right)
with $z$ for a definite value of $m$ and $m_0$. In the $Y_B~vs~z$ plot $Y_B$ freezes at a value $8.67\times10^{-11}$
(which is just midway between the the experimental bound).}
\label{etl1}
\end{figure}
 \end{center}

\begin{figure}[!h]
\includegraphics[width=5cm,height=5cm,angle=0]{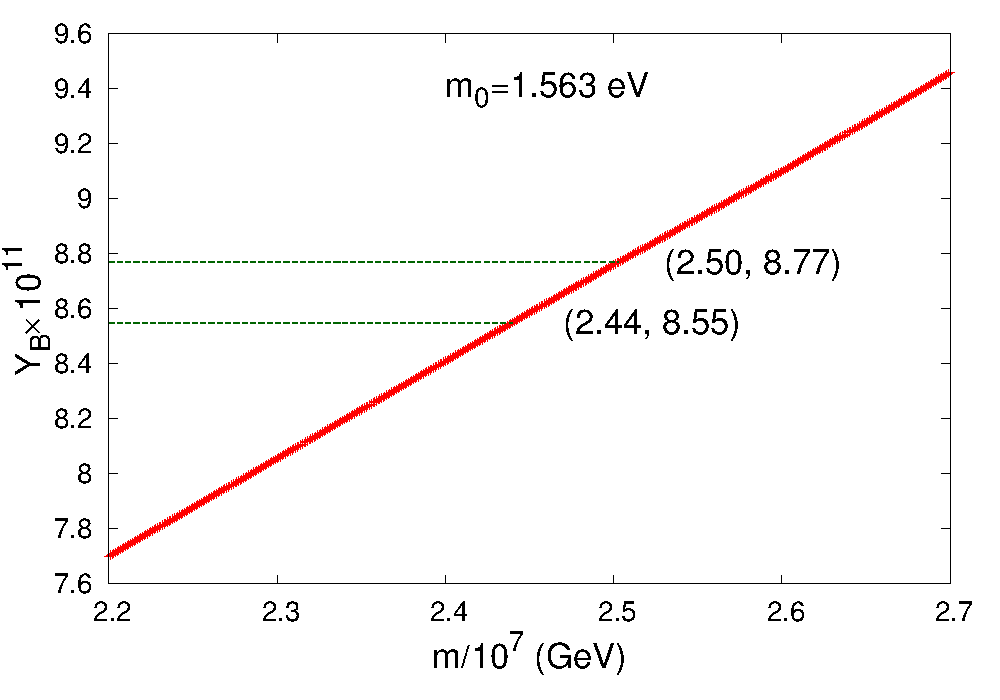}
\includegraphics[width=5cm,height=5cm,angle=0]{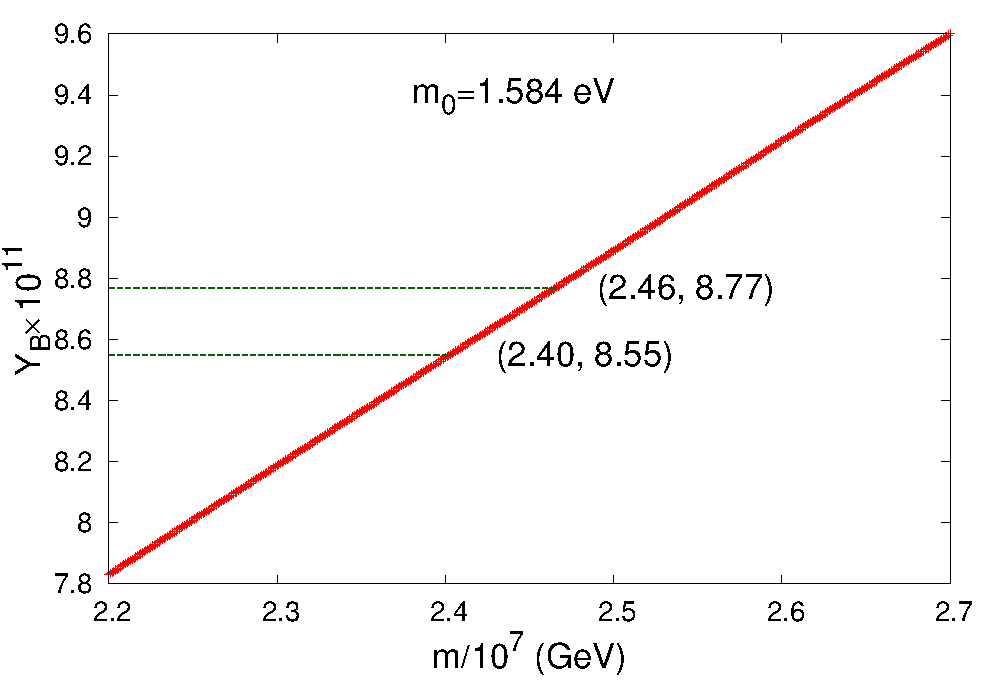}
\includegraphics[width=5cm,height=5cm,angle=0]{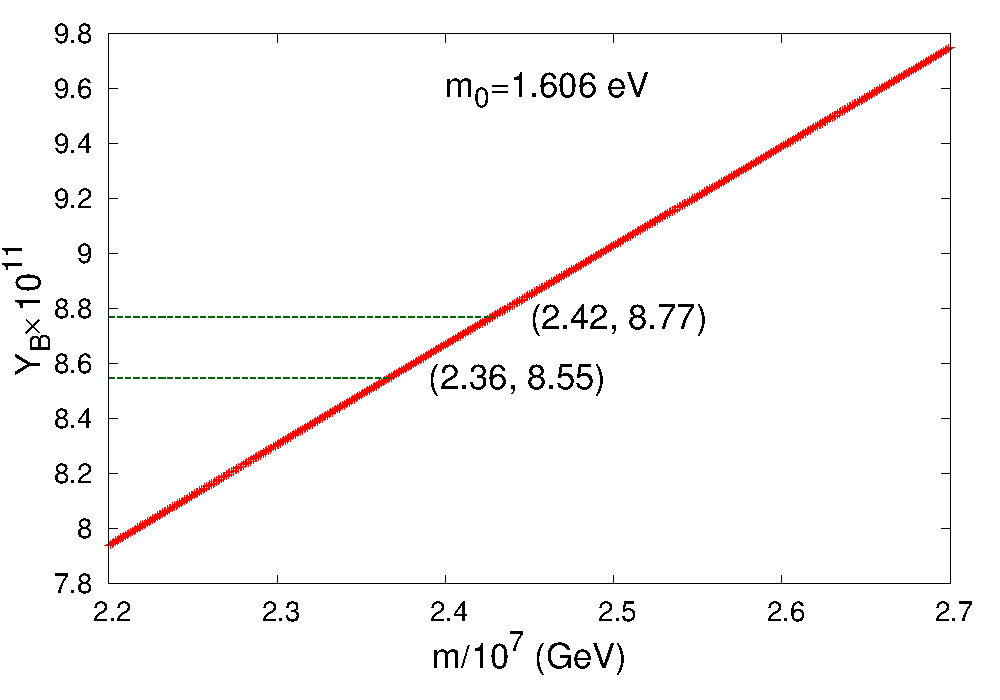}\\
\includegraphics[width=5cm,height=5cm,angle=0]{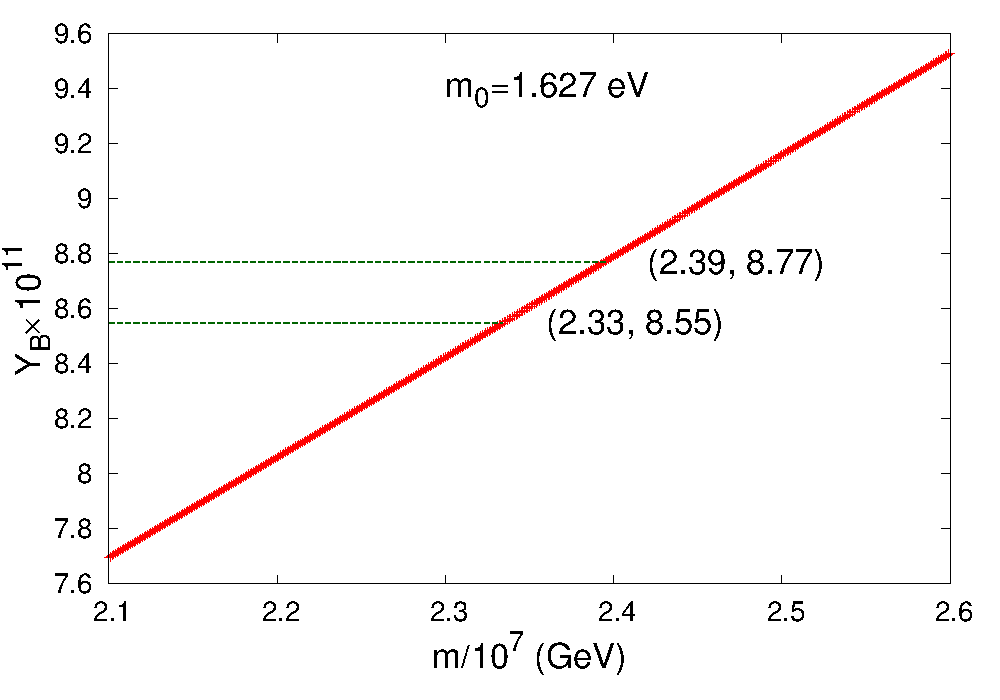}
\includegraphics[width=5cm,height=5cm,angle=0]{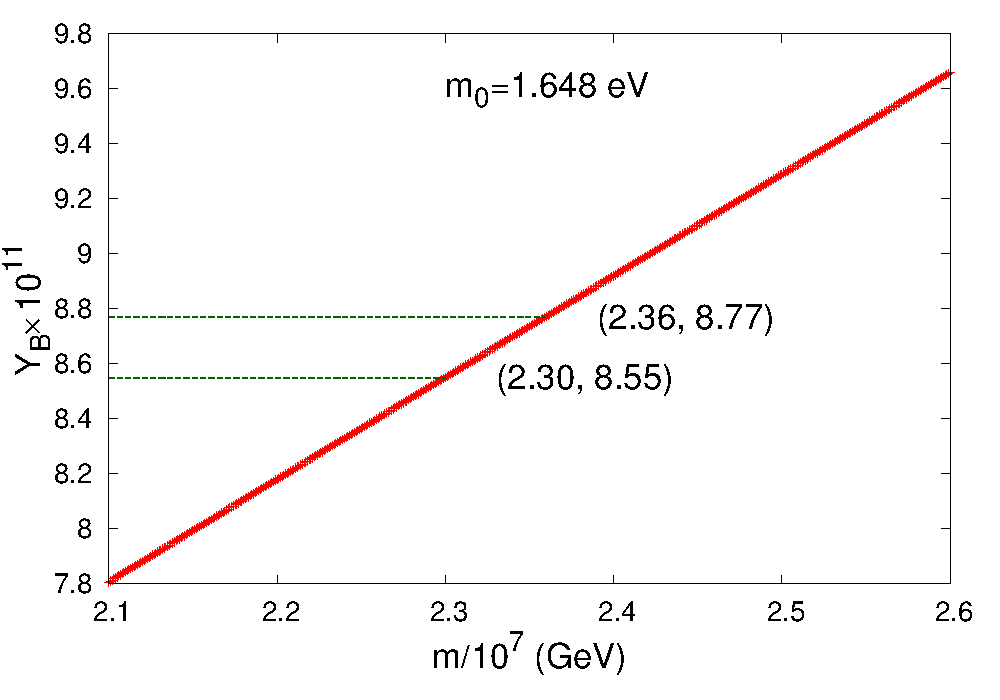}
\caption{(colour online) Plot of Final $Y_B$ with $m$ for different values of $m_0$}
\label{etab_m}
\end{figure}

To solve the sign ambiguity
of the phases and to restrict the right handed neutrino mass scale $m$  we carry out the required calculation for 
baryogenesis through leptogenesis in three different energy regimes using these set of $10$ values only.
We have one free parameter in hand, i.e the mass of lowest right handed neutrino which is chosen according
to the energy regime we are working in.
\begin{table}[!h]
\caption{Sets of parameters allowed by oscillation data for case(i) of symmetry breaking with $ \epsilon=-0.004$} \label{case1_osc}
\begin{center}
\begin{tabular}{p{1cm}|p{1cm}|p{1cm}|p{1.5cm}|p{1.5cm}|p{1.5cm}|p{2cm}|}
\cline{2-7}
 & \multicolumn{6}{c|}{parameters}\\
\hline
\multicolumn{1}{ |c| }{sets} &  $p$ & $q$ & $\alpha$ (deg.) & $\beta$ (deg.)&  $m_0\times10^9$ (GeV)& $\frac{m}{10^7}$ (GeV)\\ 
\cline{1-7}
\multicolumn{1}{ |c| }{1 } &  $0.97$  & $0.89$  & $-113.5$  & $120.5$ &  $1.563$  &   \\
\multicolumn{1}{ |c| }{ } &    &   &   &  &  $1.584$  &  unrestricted  \\ 
\multicolumn{1}{ |c| }{ } &    &   &   &  & $1.606$   &  for all \\ 
\multicolumn{1}{ |c| }{ } &    &   &   &  & $1.627$   &  $m_0$ values \\ 
\multicolumn{1}{ |c| }{ } &    &   &   &  & $1.648$   &   \\ 
\hline
\multicolumn{1}{ |c| }{2 } &  $0.97$  & $0.89$  & $113.5$  & -$120.5$ &  $1.563$  &    \\
\multicolumn{1}{ |c| }{ } &    &   &   &  &  $1.584$  &  unrestricted \\
\multicolumn{1}{ |c| }{ } &    &   &   &  &  $1.606$  &  for all  \\
\multicolumn{1}{ |c| }{ } &    &   &   &  &   $1.627$ &  $m_0$ values\\
\multicolumn{1}{ |c| }{ } &    &   &   &  &   $1.648$ &   \\
\hline
\end{tabular}
\end{center}
\end{table}
In the {\bf fully flavored} case the mass of the lightest right handed neutrino is less than $10^9$ Gev. 
All three lepton flavors
($e$, $\mu$, $\tau$) are separately active in this regime. 
Here we have to solve the set of flavor dependent coupled Boltzmann equations (\ref{BEN_Y1}, \ref{BEL_Y1}) for three
flavors ($e,~\mu,~\tau$) separately to get the evolution of the flavor asymmetries ($Y_{\Delta_e},~Y_{\Delta_\mu},~Y_{\Delta_\tau}$)
with $z$. The RHS of the Boltzmann equations are known in terms of the Lagrangian parameters which are already
restricted by oscillation data. After obtaining the $Y_{\Delta_\alpha}$ asymmetry parameters we have to follow the steps given
in subsection \ref{ff} to get the final value of the baryon asymmetry scaled by entropy density ($Y_B$)
\footnote{Final value of $Y_B$ means it is the frozen value of $Y_B\{=(n_B-n_{\bar{B}})/s\}$ which is its value at present epoch and it is related
to frozen value of $\eta_B\{=(n_B-n_{\bar{B}})/n_\gamma\}$ as $\eta_B=7.04 Y_B$.}.
~Evolution of $Y_B(=\frac{n_B-n_{\overline{B}}}{s})$ with $z$ is computed with each set of values of the parameters
$\{p,~q,~\alpha,~\beta,~m_0\}$ for different values of right handed neutrino mass $m$. It is found that while
using the first five sets ( i.e set 1 of $\{ p,~q,~\alpha,~\beta \}$ with five different $m_0$) $Y_B$ attains a
constant positive value at a high $z$ whereas $Y_B$ freezes at negative value when the calculation is done with next five sets
(i.e set 2 of $\{ p,~q,~\alpha,~\beta \}$ with five different $m_0$). Therefore set 2 of $\{ p,~q,~\alpha,~\beta \}$
can readily be discarded since experimental observations have confirmed the fact that baryon asymmetry at present 
epoch must be positive. For the first five sets final value of $Y_B$ is calculated (for different right handed neutrino 
masses) and tabulated below (Table \ref{final_bary1}). 
\begin{table}[!h]
\caption{Final value of baryon asymmetry for set 1 of $\{ p,~q,~\alpha,~\beta \}$ ($m$ and $m_0$ are in GeV)} \label{final_bary1}
\begin{center}
\begin{tabular}{p{2cm}|p{1cm}|p{1cm}|p{1cm}|p{1cm}|p{1cm}|p{1cm}|p{1cm}|p{1cm}|}
\cline{2-9}
 & \multicolumn{8}{c|}{\bf $Y_B\times10^{11}$ }\\
\hline
\multicolumn{1}{ |c| }{\backslashbox{$m_0\times10^9$}{$\frac{m}{10^7}$}} &  $2.0$ & $2.2$ & $2.4$ & $2.5$ &$2.6$ & $2.7$ & $2.9$ & $3.0$ \\ \cline{1-9} 
\multicolumn{1}{ |c| }{$1.563$ } & $7.00$ & $7.70$ & $8.41$ & $8.76$ & $9.10$ & $9.46$ & $10.16$ & $10.51$ \\
\hline
\multicolumn{1}{ |c| }{$1.584$ } & $7.11$ & $7.83$ & $8.54$ & $8.89$ & $9.25$ & $9.60$ & $10.31$ &  $10.67$ \\
\hline
\multicolumn{1}{ |c| }{ $1.606$} & $7.22$ & $7.94$ & $8.67$ & $9.03$ & $9.39$ & $9.75$ & $10.47$ & $10.83$\\
\hline
\multicolumn{1}{ |c| }{$1.627$ } & $7.33$ & $8.06$ & $8.79$ & $9.16$ & $9.53$ & $9.89$ & $10.63$ & $10.99$\\
\hline
\multicolumn{1}{ |c| }{ $1.648$} & $7.43$ & $8.18$ & $8.92$ & $9.29$ & $9.66$ & $10.03$ & $10.78$ &  $11.15$\\
\hline
\end{tabular}
\end{center}
\end{table}
As an example we pick the case $(m=2.4\times10^7~ {\rm GeV},~m_0=1.606\times10^{-9}~{\rm GeV})$ (for which $Y_B$ is within
the experimental range) and show the variation of flavor asymmetries ($Y_{\Delta_\alpha}$) and 
baryon asymmetry ($Y_B$) with $z$ in Fig.\ref{etl1}.
The sign of different asymmetries ($Y_{\Delta_e},~Y_{\Delta_\mu},~Y_{\Delta_\tau}$) for various values of z are 
shown in the Table \ref{asy}. An important observation of our numerical estimation should also be mentioned in 
this context that $|Y_{\Delta_\tau}|$ is always greater than $|Y_{\Delta_e}+Y_{\Delta_\mu}|$. It is clear from 
Table \ref{asy} that same quantities in set 1 and set 2 bears a relative opposite sign
\footnote{Here set 1 corresponds to $\{p,q,\alpha,\beta,m_0\}$ and set 2 corresponds to $\{p,q,-\alpha,-\beta,m_0\}$.
Examining expression of the flavor dependent CP asymmetry parameter $\varepsilon^\alpha_i$ (eq.(\ref{epsi}))
we find that apart from $p,q,m_0$ it contains $sine$ functions of $\alpha$, $\beta$ and $(\alpha \pm \beta)$. Therefore
for example if  $\varepsilon^\alpha_i(p,q,\alpha,\beta,m_0)$ is positive, $\varepsilon^\alpha_i(p,q,-\alpha,-\beta,m_0)$
must be negative. These $\varepsilon^\alpha_i$ s are then inserted in the Boltzmann equation (eq.(\ref{BEL_Y1})) to find the flavor
dependent $Y_{\Delta_\alpha}$ asymmetries. This is why the relative opposite sign appears between the same
quantities of set 1 and set 2.
}. Here we are showing the plots for set 1 only, which survives the baryon asymmetry bound.
\begin{table}[!h]
\caption{signs of different asymmetries at different $z$ values} \label{asy}
\begin{center}
\begin{tabular}{p{2cm}|p{.8cm}p{.8cm}p{.8cm}p{.8cm}|p{.8cm}p{.8cm}p{.8cm}p{.8cm}|p{.8cm}p{.8cm}p{.8cm}p{.8cm}|}
\cline{2-13}
 &\multicolumn{4}{c|}{$z=0.01\rightarrow 0.02$}& \multicolumn{4}{c|}{$z=0.02\rightarrow 0.5$}&\multicolumn{4}{c|}{$z>0.5$}\\
\hline
\multicolumn{1}{ |c| }{}& $Y_{\Delta_e}$  & $Y_{\Delta_\mu}$ & $Y_{\Delta_\tau}$ & $Y_B$  & $Y_{\Delta_e}$  & $Y_{\Delta_\mu}$ & $Y_{\Delta_\tau}$ & $Y_B$ &$Y_{\Delta_e}$ & $Y_{\Delta_\mu}$ & $Y_{\Delta_\tau}$ & $Y_B$ \\ \cline{1-13} 
\multicolumn{1}{ |c| }{set 1 } & -ve &-ve &+ve & +ve & +ve  & +ve & -ve & -ve & -ve & -ve & +ve & +ve\\
\multicolumn{1}{ |c| }{set 2 } & +ve & +ve & -ve & -ve & -ve & -ve & +ve & +ve & +ve & +ve  & -ve & -ve \\
\hline
\end{tabular}
\end{center}
\end{table}
From Table \ref{final_bary1} it is clear that we can get a bound on the right handed neutrino mass ($m$)
using the $Y_B$ constraint $(8.55<Y_B\times10^{11}<8.77)$\footnote{or equivalently the bound on $\eta_B$ is
$(6.02<\eta_B\times10^{10}<6.18)$\cite{Dev:2015uca}.} for each value of $m_0$ (rather for each 
value of the set $\{p,~q,~\alpha,~\beta,~m_0\}$). Similarly we plot (Fig.\ref{etab_m}) the final value of 
$Y_B$ with the mass of the lowest right handed neutrino ($m$) for five different values of $m_0$.
In each of these plots we draw two lines parallel to abscissa, one at $Y_B=8.55\times10^{-11}$ and
the other at $Y_B=8.77\times10^{-11}$. The value of $m$ where the lines meet the $Y_B~vs~m$ curve 
give the lower and upper bound on $m$ respectively. Allowed range of $m$ for different $m_0$ s are tabulated
in Table \ref{range_m1}. It is to be noticed that after imposition of baryon asymmetry bound the sign of the phase 
parameters and the the mass of the right handed neutrinos get a restriction and
the fully constrained (by both oscillation data and baryon asymmetry bound) 
parameter space is presented in Table \ref{full_case1}.
\begin{table}[!h]
\caption{Range of $m$ allowed by $Y_B$ constraint for different $m_0$ values} \label{range_m1}
\begin{center}
\begin{tabular}{|p{2.6cm}|p{2cm}|p{2cm}|p{2cm}|p{2cm}|p{2cm}|}
\hline
$m_0\times10^9$ (GeV) & $1.563$ & $1.584$ & $1.606$ & $1.627$ & $1.648$ \\
\hline
$\frac{m}{10^7}$ (GeV) & $2.44-2.50$ & $2.40-2.46$& $2.36-2.42$ & $2.33-2.39$ & $2.30-2.36$\\
\hline 
\end{tabular}
\end{center}
\end{table}
\begin{table}[!h]
\caption{Sets of parameters allowed by both oscillation data and baryon asymmetry bound for case(i) of symmetry breaking 
with $ \epsilon=-0.004$ (In fully flavored regime)} \label{full_case1}
\begin{center}
\begin{tabular}{p{1cm}|p{1cm}|p{1cm}|p{1.5cm}|p{1.5cm}|p{1.5cm}|p{2.3cm}|}
\cline{2-7}
 & \multicolumn{6}{c|}{parameters}\\
\hline
\multicolumn{1}{ |c| }{sets} &  $p$ & $q$ & $\alpha$ (deg.) & $\beta$ (deg.)&  $m_0\times10^9$ (GeV)& $\frac{m}{10^7}$ (GeV)\\ 
\cline{1-7}
\multicolumn{1}{ |c| }{1 } &  $0.97$  & $0.89$  & $-113.5$  & $120.5$ &  $1.563$  & $2.44-2.50$  \\
\multicolumn{1}{ |c| }{ } &    &   &   &  &  $1.584$  & $2.40-2.46$   \\ 
\multicolumn{1}{ |c| }{ } &    &   &   &  & $1.606$   &  $2.36-2.42$ \\ 
\multicolumn{1}{ |c| }{ } &    &   &   &  & $1.627$   &  $2.33-2.39$ \\ 
\multicolumn{1}{ |c| }{ } &    &   &   &  & $1.648$   &   $2.30-2.36$ \\ 
\hline
\end{tabular}
\end{center}
\end{table}

In the {\bf $\tau$-flavored} case the mass of the lightest right handed neutrino is less
than $10^{12}$ Gev but greater than $10^9$ GeV.
In this regime we can not distinguish between $e$ and $\mu$ flavors, whereas the $\tau$ flavor 
is decoupled. In this case the Boltzmann equations (\ref{BEN_Y1}, \ref{BEL_Y1}) have to be solved for 
two flavors only, first one is for combined effect of $e$, $\mu$ and the second one is for $\tau$.
To find the combined asymmetry $Y_{\Delta_2}(=Y_{\Delta_e}+Y_{\Delta_\mu})$ we have to replace the $\alpha$ dependent 
terms in the RHS of eq.(\ref{BEL_Y1}) by the sum over $2$ flavors $e$, $\mu$ and $Y_{\Delta_\tau}$ can 
be obtained simply by solving eq.(\ref{BEL_Y1}) for $\tau$ flavor. 
Once we get $Y_{\Delta_2}(z)$ and $Y_{\Delta_\tau}(z)$ the frozen value of the baryon asymmetry scaled by entropy density
$Y_B$ can be calculated easily following the steps given in subsection \ref{pf}. 
In this case too we compute variation of $Y_B$ with $z$ for ten different values of 
the set  $\{p,~q,~\alpha,~\beta,~m_0 \}$ (Table \ref{case1_osc}) and the set 2 of  $\{p,~q,~\alpha,~\beta \}$
is discarded due to the same argument as discussed in the {\bf fully flavored} case.
Final value of the ratio of baryon asymmetry to entropy density ($Y_B$) is calculated using set 1 of $\{p,~q,~\alpha,~\beta\}$
with five different values of $m_0$ for various values of right handed neutrino mass ($m$) in 
the range ($10^9~{\rm GeV}<m<10^{12}~{\rm GeV}$). It is found that $Y_B$ in each of the combinations
is far above the experimental upper bound.

Finally for the {\bf unflavored} case the mass of the lightest right handed neutrino is greater than $10^{12}$ GeV and we can not differentiate
between $e$, $\mu$ and $\tau$ flavors. For this case the Boltzmann equation for lepton flavor asymmetry is flavor 
independent. The flavor index dependent quantities in the RHS of the Boltzmann equation are replaced by the sum over 
three flavors ($e,~\mu,~\tau$). Solving the Boltzmann equation we get $Y_\Delta(z)$ from which the baryon asymmetry
parameter ($Y_B$) can be computed using the formulas given in subsection \ref{uf}. 
In spite of maximum value of CP asymmetry parameters in this regime the final value of lepton flavor asymmetry turns out to be 
zero. The reason behind this anomalous behaviour is discussed  below.

The cyclic symmetry invariant $m_D$ and $M_R=diag(m,~m,~m+\epsilon)$ 
~dictates the CP asymmetry parameters ($\varepsilon_i=\sum\limits_{\alpha}\varepsilon^\alpha_i$) as
$\varepsilon_1=-\varepsilon_2$ and $\varepsilon_3=0$ (detailed calculation shown in appendix \ref{a1}). 
Now looking at second Boltzmann equation (\ref{BEL_Y1}) it is 
clear that at the starting point of iteration the second term in the RHS is zero since we have taken the 
initial condition \footnote{It is to be noted that second term of eq.(\ref{BEL_Y1}) is basically the washout term which tends to 
erase any pre-existing asymmetry. Thus even if we start with some pre-existing asymmetry as initial
condition this second term will have no effect other than diminishing that asymmetry.}$Y_\Delta(z=0)=0$.
So $\frac{d Y_\Delta}{dz}$ consists of sum of three terms involving 
$\varepsilon_1$, $\varepsilon_2$ and $\varepsilon_3$ respectively. The term containing $\varepsilon_3$ has
null contribution whereas the terms involving $\varepsilon_1$ and $\varepsilon_2$ cancels each 
other\footnote{Coefficients of $\varepsilon_1$ and $\varepsilon_2$ differ only by the parameter $m_{N_i}$ (mass of the 
right handed neutrino of corresponding generation), i.e apart from identical common factors $\varepsilon_1$ contains
$m_{N_1}$ whereas that of $\varepsilon_2$ contains $m_{N_2}$. But in our breaking scheme 
$(M_R=diag(m,~m,~m+\epsilon))~\Rightarrow m_{N_1}=m_{N_2}$. Therefore the terms involving 
$\varepsilon_1$ and $\varepsilon_2$ in the second Boltzmann equation are exactly same and thus cancels 
each other when $\varepsilon_1=-\varepsilon_2$. 
} 
as a result we get $\frac{d Y_\Delta}{dz}=0$. So generation of lepton asymmetry as well as baryon asymmetry 
is not possible in this unflavored leptogenesis scenario.

As a result of imposing baryon asymmetry bound together with neutrino oscillation data,
the right handed neutrino mass scale is restricted and among set 1 and set 2, only the parameters
belonging to set 1 are selected. Hence the value of the Dirac CP phase  $\delta_{\rm CP}$ is fixed to 
$29.38^\circ$.
 
\subsection{Numerical analysis for case(ii) of symmetry breaking }
As mentioned in eq.(\ref{3cat_brk}) for this case the symmetry breaking parameter is introduced in the \textquoteleft$22$\textquoteright~
element of $M_R$. The mixing angles and mass eigenvalues are calculated using the resulting mass matrix given in eq.(\ref{mnu2}).
In this case the lowest allowed value
of the breaking parameter $\epsilon$ is $-0.004$ and for this value of $\epsilon$ we get $16$ allowed
values of the parameter set $\{p,~q,~\alpha,~\beta,~m_0\}$.
We have six different sets of $\{p,~q,~\alpha,~\beta\}$ and for set 1 and set 2 of $\{p,~q,~\alpha,~\beta\}$
there are six different $m_0$ values whereas set 3-6 has one allowed $m_0$ value each. Thus in total
we have $16(=6+6+1+1+1+1)$ values for the set of parameters $\{p,~q,~\alpha,~\beta,~m_0\}$ which are
tabulated in Table \ref{case2}. In this case (case(ii)) too
normal hierarchy is preferred but $\theta_{23}$ is now selected in the 2nd octant ($48.07^\circ-49.09^\circ$).
The sign of $\alpha$ and $\beta$  remain unsettled between the following sets: (1 and 2), (3 and 4), (5 and 6). 
These in effect again produce sign ambiguity in Dirac CP phase: 
 $\delta_{\rm CP}=44.45^\circ$ for the set 1 with ($\alpha=-125^\circ$, $\beta=116^\circ$)  and
 $\delta_{\rm CP}=-44.45^\circ$ for the set 2 with ($\alpha=125^\circ$, $\beta=-116^\circ$),
$\delta_{\rm CP}=16.56^\circ$ for the set 3 with ($\alpha=-121^\circ$, $\beta=113.5^\circ$)  and
 $\delta_{\rm CP}=-16.56^\circ$ for the set 4 with ($\alpha=121^\circ$, $\beta=-113.5^\circ$),
$\delta_{\rm CP}=14.27^\circ$ for the set 5 with ($\alpha=-117.5^\circ$, $\beta=116^\circ$)  and
 $\delta_{\rm CP}=-14.27^\circ$ for the set 6 with ($\alpha=117.5^\circ$, $\beta=-116^\circ$).
To solve the sign ambiguity
of the phases and to restrict the right handed neutrino mass scale $m$  for this case also, we investigate all three subcases of leptogenesis 
namely, fully flavored, $\tau$-flavored and unflavored. 
\begin{table}[!h]
\caption{Sets of parameters allowed by oscillation data for case(ii) of symmetry breaking with $ \epsilon=-0.004$} \label{case2}
\begin{center}
\begin{tabular}{p{1cm}|p{1cm}|p{1cm}|p{1.5cm}|p{1.5cm}|p{1.5cm}|p{2cm}|}
\cline{2-7}
 & \multicolumn{6}{c|}{parameters}\\
\hline
\multicolumn{1}{ |c| }{sets} &  $p$ & $q$ & $\alpha$ (deg.) & $\beta$ (deg.)&  $m_0\times10^9$ (GeV)& $\frac{m}{10^7}$ (GeV)\\ 
\cline{1-7}
\multicolumn{1}{ |c| }{1 } &  $0.97$  & $1.05$  & $-125$  & $116$ &  $1.881$  &   \\
\multicolumn{1}{ |c| }{ } &    &   &   &  &  $1.907$  &  unrestricted  \\ 
\multicolumn{1}{ |c| }{ } &    &   &   &  & $1.933$   &  for all \\ 
\multicolumn{1}{ |c| }{ } &    &   &   &  & $1.959$   &  $m_0$ values \\ 
\multicolumn{1}{ |c| }{ } &    &   &   &  & $1.984$   &   \\ 
\multicolumn{1}{ |c| }{ } &    &   &   &  & $2.009$   &   \\
\hline
\multicolumn{1}{ |c| }{2 } &  $0.97$  & $1.05$  & $125$  & $-116$ &  $1.881$  &    \\
\multicolumn{1}{ |c| }{ } &    &   &   &  &  $1.907$  &  unrestricted \\
\multicolumn{1}{ |c| }{ } &    &   &   &  &  $1.933$  &  for all  \\
\multicolumn{1}{ |c| }{ } &    &   &   &  &   $1.959$ &  $m_0$ values\\
\multicolumn{1}{ |c| }{ } &    &   &   &  &   $1.984$ &   \\
\multicolumn{1}{ |c| }{ } &    &   &   &  &   $2.009$ &   \\
\hline
\multicolumn{1}{ |c| }{3 } &  $0.89$  & $0.95$  & $-121$  & $113.5$ &  $1.717$  &  unrestricted  \\
\hline
\multicolumn{1}{ |c| }{4 } &  $0.89$  & $0.95$  & $121$  & $-113.5$ &  $1.717$  &  unrestricted  \\
\hline
\multicolumn{1}{ |c| }{5 } &  $0.91$  & $0.91$  & $-117.5$  & $116$ &  $1.683$  &  unrestricted  \\
\hline
\multicolumn{1}{ |c| }{6 } &  $0.91$  & $0.91$  & $117.5$  & $-116$ &  $1.683$  &  unrestricted  \\
\hline
\end{tabular}
\end{center}
\end{table}

\begin{center}
\begin{figure}[!h]
\includegraphics[width=7.5cm,height=6.5cm,angle=0]{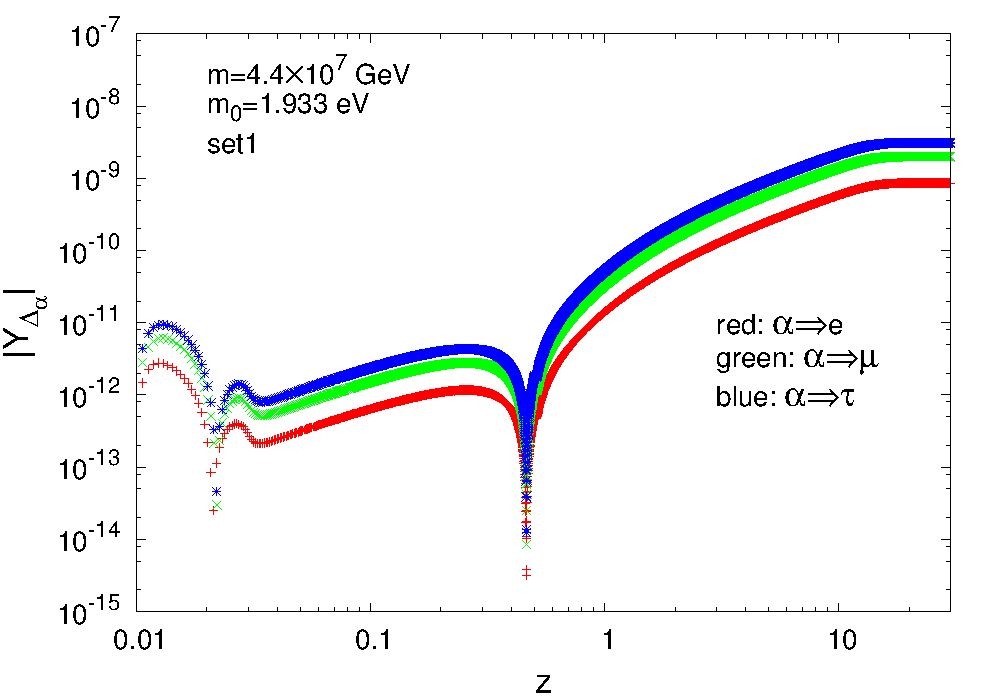}
\hspace{1.0cm}
\includegraphics[width=7.5cm,height=6.5cm,angle=0]{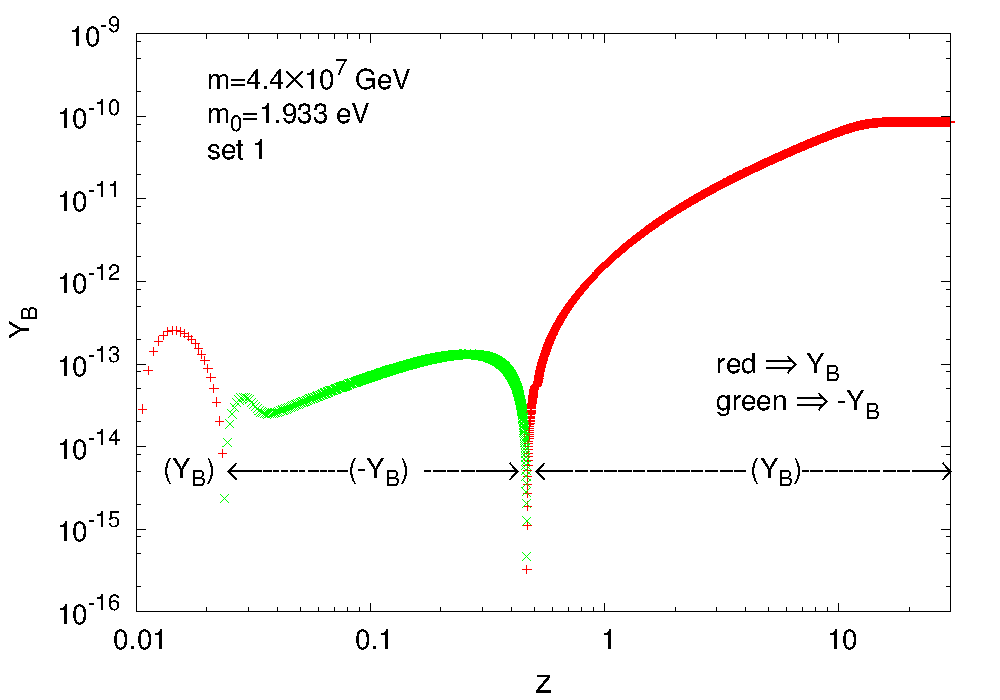}
\caption{(colour online) Plot of flavor asymmetries($Y_{\Delta_\alpha}$) (left), baryon asymmetry($Y_B$) (right)
with $z$ for a definite value of $m$ and $m_0$. In $Y_B~vs~z$ plot $Y_B$ freezes at a value $8.60\times10^{-11}$.}
\label{etl2}
\end{figure}
 \end{center}

\begin{figure}[!h]
\includegraphics[width=5cm,height=5cm,angle=0]{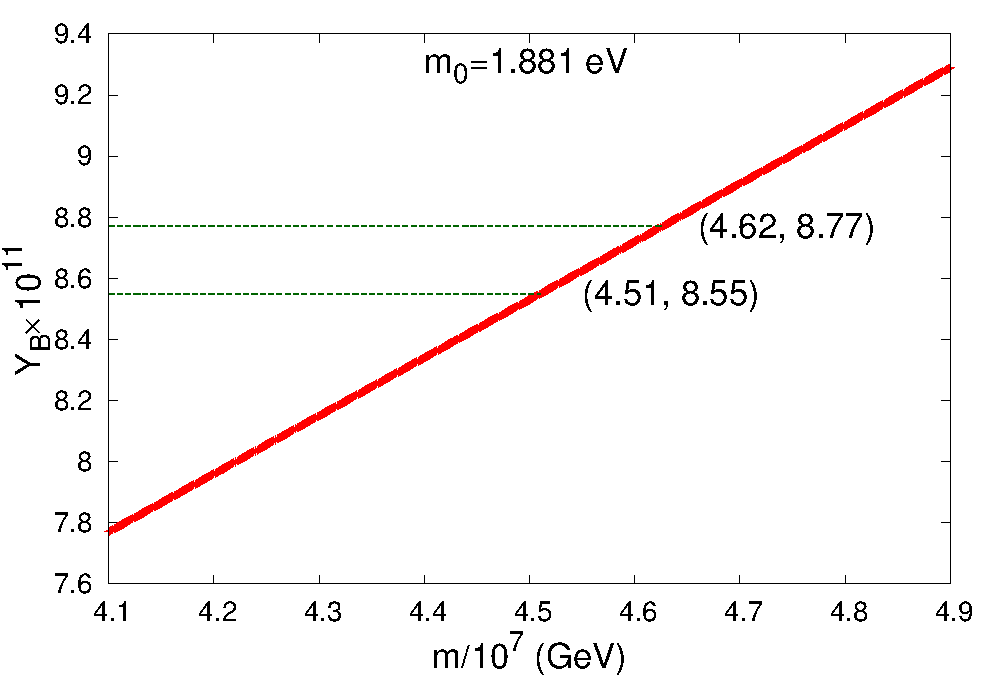}
\includegraphics[width=5cm,height=5cm,angle=0]{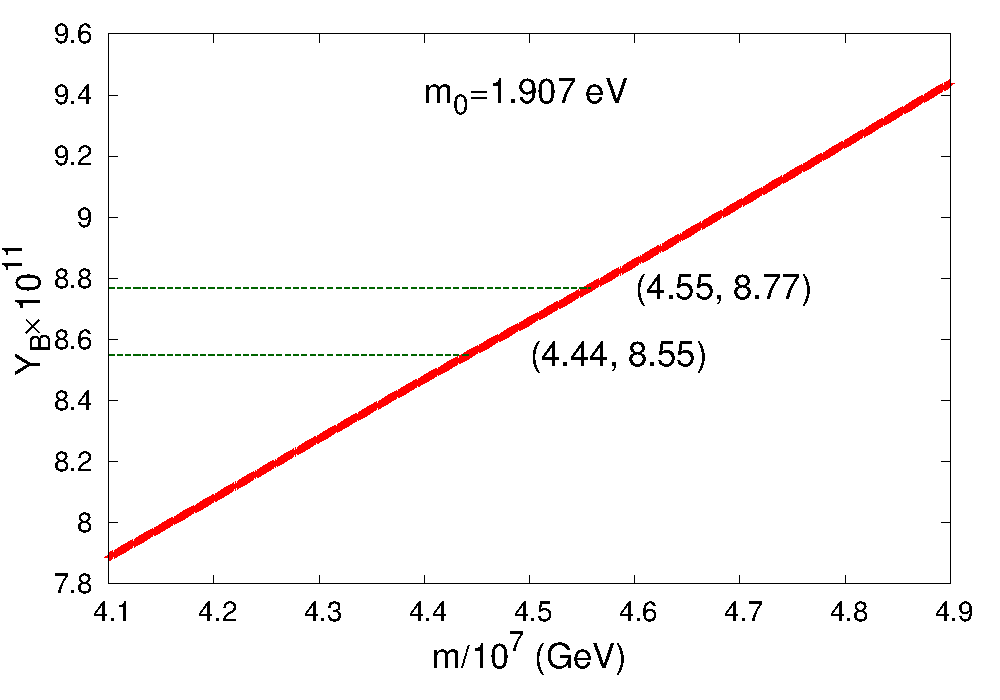}
\includegraphics[width=5cm,height=5cm,angle=0]{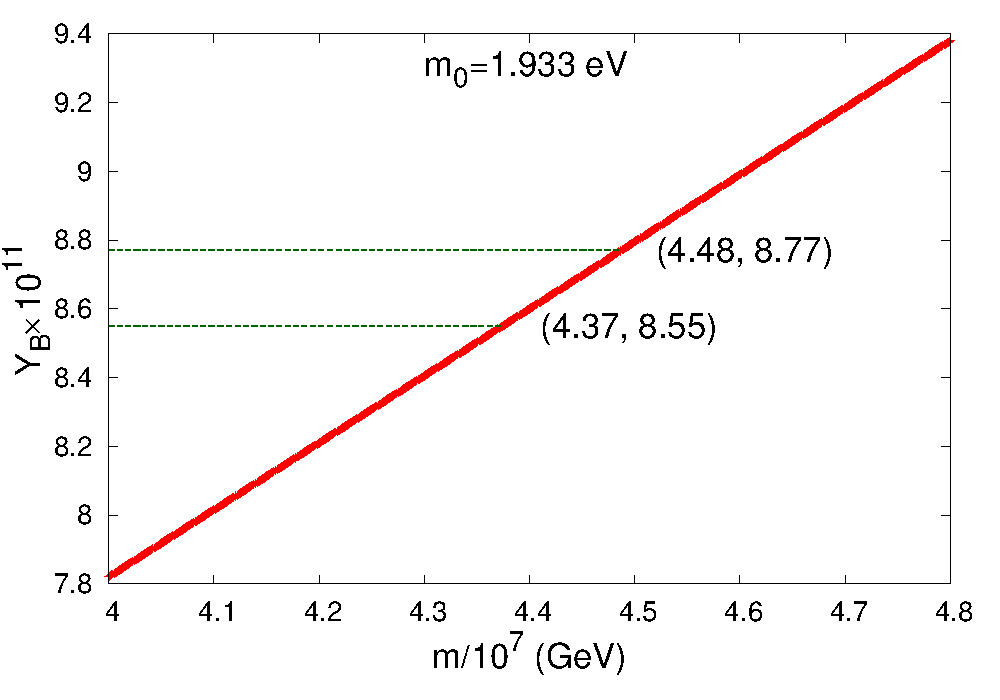}\\
\includegraphics[width=5cm,height=5cm,angle=0]{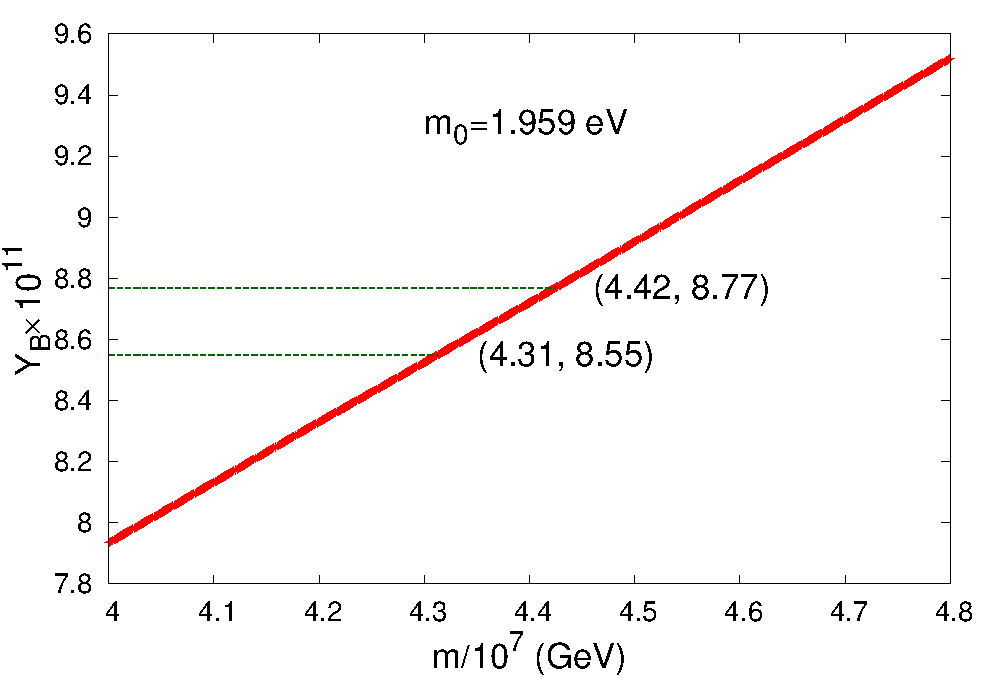}
\includegraphics[width=5cm,height=5cm,angle=0]{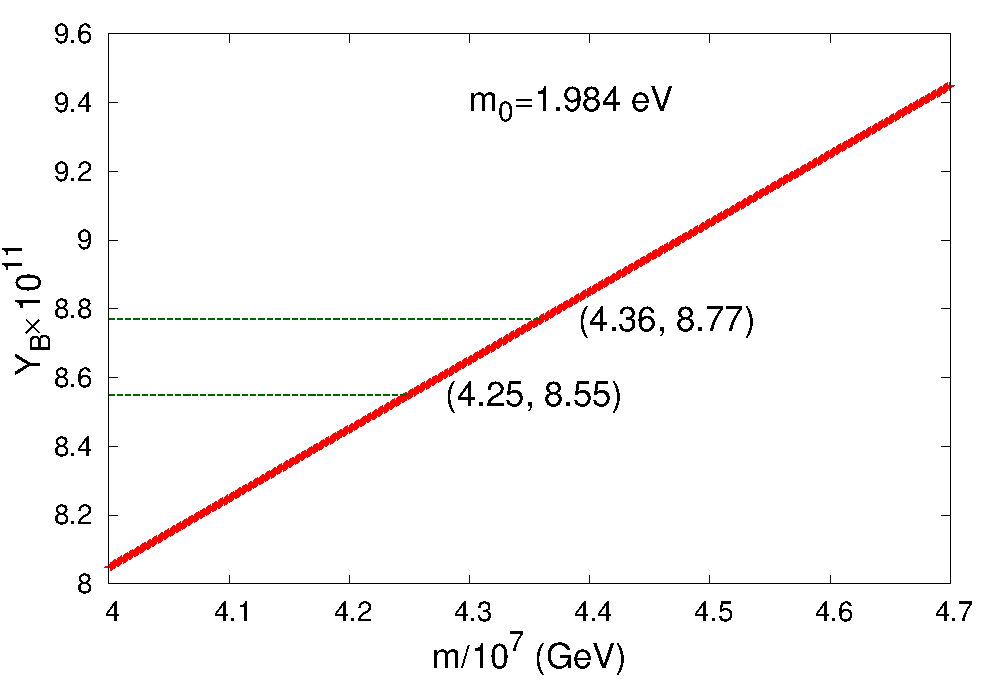}
\includegraphics[width=5cm,height=5cm,angle=0]{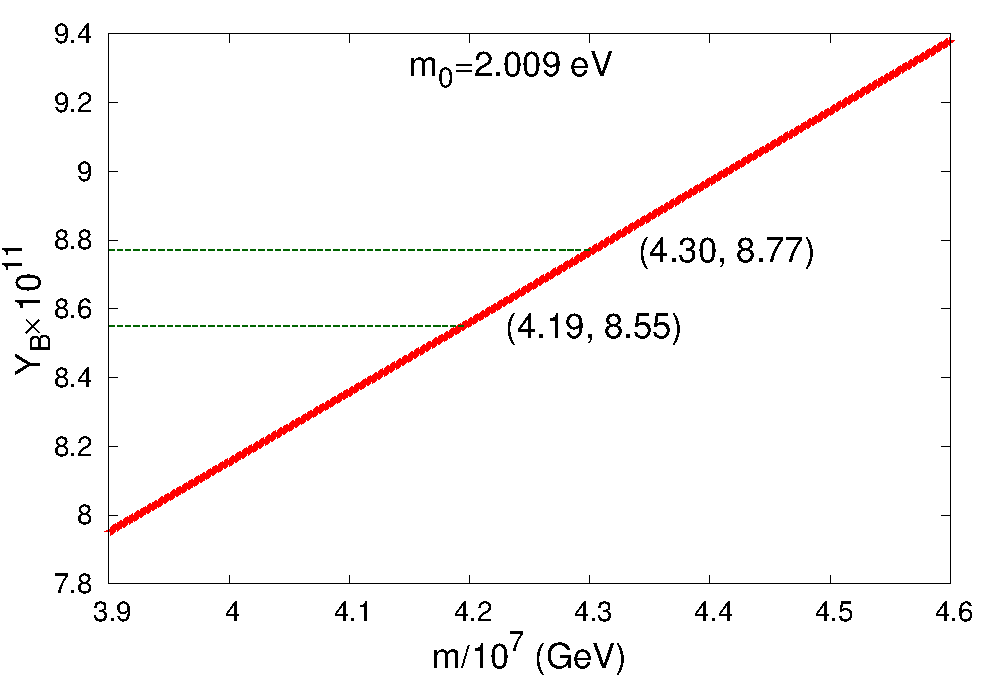}
\caption{(colour online) Plot of Final $Y_B$ with $m$ for different values of $m_0$ with set 1 of $\{p,~q,~\alpha,~\beta\}$}
\label{etab_m2}
\end{figure}

\begin{figure}[!h]
\begin{center}
\includegraphics[width=5cm,height=5cm,angle=0]{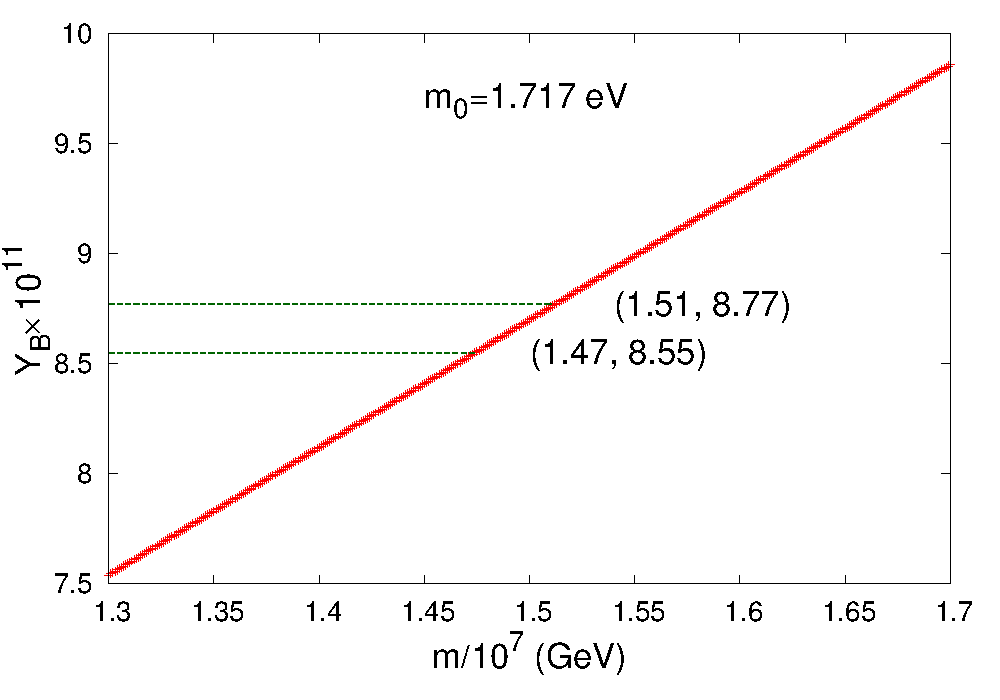}
\hspace{1cm}
\includegraphics[width=5cm,height=5cm,angle=0]{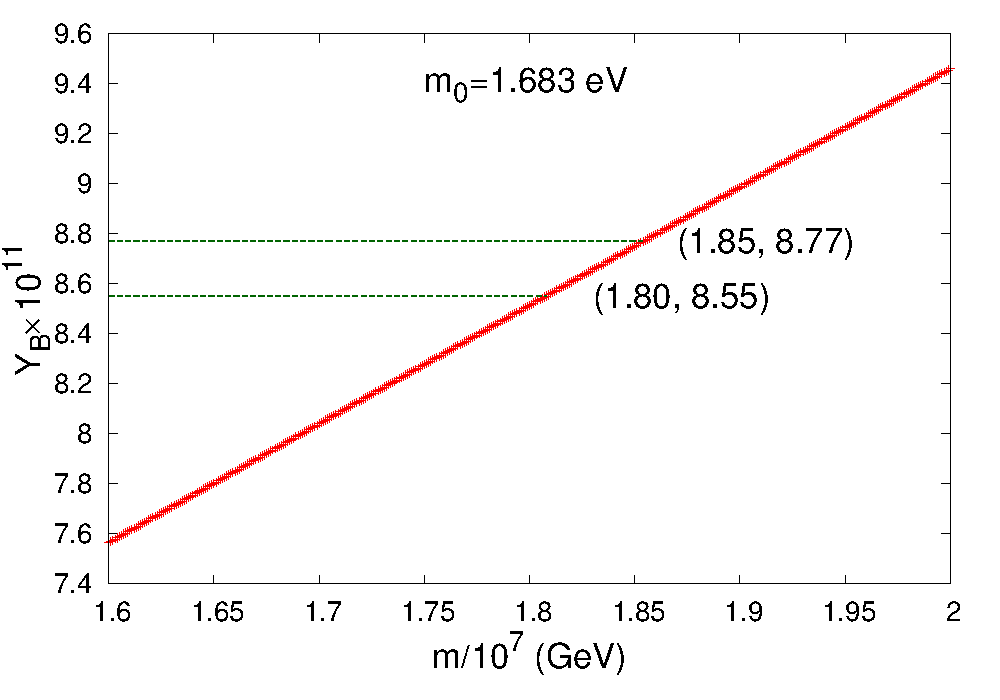}
\caption{(colour online) Plot of Final $Y_B$ with $m$ for different values of $m_0$ with set 3 (left) and set 5 (right) 
of $\{p,~q,~\alpha,~\beta\}$}
\label{etab_m3}
\end{center}
\end{figure}

\paragraph{}
Following the same steps as the previous case (case(i) of symmetry breaking) in {\bf fully flavored} regime
evolution of $Y_B$ with $z$ is computed with each of the $16$ values of the set
$\{p,~q,~\alpha,~\beta,~m_0\}$ for different values of right handed neutrino mass $m$.
After the numerical analysis of each of the above combinations it is found that
for set 2, 4, 6 of $\{p,~q,~\alpha,~\beta \}$
$Y_B$ produced at high $z$ value attain a fixed negative value and
thus these three sets can be discarded. Therefore we are left
with set 1, 3, 5 of $\{p,~q,~\alpha,~\beta \}$ and their corresponding $m_0$
values (six for set 1 and one each for set 3 and set 5) which gives rise to constant positive
$Y_B$ at high $z$. The final value of ratio of baryon asymmetry to entropy density ($Y_B$)
for each set of values with different $m$ values are presented in Table \ref{final_bary2}.
As an example we choose the case $(m=4.4\times10^7~{\rm GeV},~m_0=1.933\times10^{-9}~{\rm GeV})$ of
set 1 and show the evolution of flavor asymmetries ($Y_{\Delta_\alpha}$) and baryon asymmetry ($Y_B$) with $z$ in Fig.\ref{etl2}.
The sign of the different flavor asymmetry parameters at various $z$ values are shown in 
Table \ref{asy2}\footnote{It is to be noted that the table is for set 1 only which survives the baryon asymmetry bound. The 
sign of the corresponding parameters belonging to set 2 (complex conjugate of set 1) will have a relative negative sign.}
and in this case too the numerical analysis reveals that $|Y_{\Delta_\tau}|>|Y_{\Delta_e}+Y_{\Delta_\mu}|$.
\begin{table}[!h]
\caption{Final value of baryon asymmetry for set 1, 3, 5 of $\{ p,~q,~\alpha,~\beta \}$ ($m$ and $m_0$ are in GeV)} \label{final_bary2}
\begin{center}
\begin{tabular}{p{2cm}|p{1cm}|p{1cm}|p{1cm}|p{1cm}|p{1cm}|p{1cm}|p{1cm}|p{1cm}|p{1cm}|}
\cline{2-9}
 & \multicolumn{8}{c|}{\bf $Y_B\times10^{11}$ }\\
\cline{1-9}
\multicolumn{1}{ |c| }{\backslashbox{$m_0\times10^9$}{$\frac{m}{10^7}$}} & $1.3$  & $1.5$ & $1.7$ & $2.0$ & $4.2$ & $4.4$ & $4.6$ & $4.8$ \\ \cline{1-10} 
\multicolumn{1}{ |c| }{ $1.881$ } & $2.46$ & $2.84$ & $3.22$ & $3.79$ & $7.96$  & $8.34$ & $8.72$ & $9.10$ &\\
\cline{1-9}
\multicolumn{1}{ |c| }{$1.907$ } & $2.50$ & $2.89$ & $3.27$ & $3.85$ & $8.08$ & $8.47$ & $8.85$ & $9.24$  &\\
\cline{1-9}
\multicolumn{1}{ |c| }{ $1.933$} & $2.54$ & $2.93$ & $3.32$ & $3.91$ & $8.21$ & $8.60$ & $8.99$ & $9.38$ & set 1\\
\cline{1-9}
\multicolumn{1}{ |c| }{$1.959$ } & $2.57$ & $2.97$ & $3.37$ & $3.96$ & $8.33$ & $8.72$ & $9.12$ & $9.52$ &\\
\cline{1-9}
\multicolumn{1}{ |c| }{$1.984$ } & $2.61$ & $3.01$ & $3.42$ & $4.02$ & $8.45$ & $8.85$ & $9.25$ & $9.65$ &\\
\cline{1-9}
\multicolumn{1}{ |c| }{$2.009$ } & $2.65$ & $3.06$ & $3.46$ & $4.08$ & $8.56$ & $8.97$ & $9.38$ & $9.79$ &\\
\hline
\multicolumn{1}{ |c| }{$1.717$ } & $7.54$ & $8.70$ & $9.86$ & $11.61$ & $24.38$ & $25.54$ & $26.70$ & $27.86$ &set 3\\
\hline
\multicolumn{1}{ |c| }{$1.683$ } & $6.15$ & $7.09$ & $8.04$ & $9.46$ & $19.87$ & $20.81$ & $21.76$ & $22.70$ &set 5\\
\hline
\end{tabular}
\end{center}
\end{table}
\begin{table}[!h]
\caption{signs of different asymmetries at different $z$ values} \label{asy2}
\begin{center}
\begin{tabular}{p{2cm}|p{.8cm}p{.8cm}p{.8cm}p{.8cm}|p{.8cm}p{.8cm}p{.8cm}p{.8cm}|p{.8cm}p{.8cm}p{.8cm}p{.8cm}|}
\cline{2-13}
 &\multicolumn{4}{c|}{$z=0.01\rightarrow 0.02$}& \multicolumn{4}{c|}{$z=0.02\rightarrow 0.46$}&\multicolumn{4}{c|}{$z>0.46$}\\
\hline
\multicolumn{1}{ |c| }{}& $Y_{\Delta_e}$  & $Y_{\Delta_\mu}$ & $Y_{\Delta_\tau}$ & $Y_B$  & $Y_{\Delta_e}$  & $Y_{\Delta_\mu}$ & $Y_{\Delta_\tau}$ & $Y_B$ &$Y_{\Delta_e}$ & $Y_{\Delta_\mu}$ & $Y_{\Delta_\tau}$ & $Y_B$ \\ \cline{1-13} 
\multicolumn{1}{ |c| }{set 1 } & -ve &-ve &+ve & +ve & +ve  & +ve & -ve & -ve & -ve & -ve & +ve & +ve\\
\hline
\end{tabular}
\end{center}
\end{table}
To get a bound on right handed neutrino mass we plot in Fig.\ref{etab_m2} and Fig.\ref{etab_m3} the final value of $Y_B$ with $m$ for different values of $m_0$
(or rather for different values of the parameter set $\{p,~q,~\alpha,~\beta,~m_0\}$) showing the allowed region and the 
range of $m$ thus obtained are presented in Table \ref{range_m2} and the finally surviving parameter space
(after imposing the baryon asymmetry bound) is shown in Table \ref{full_case2}.
\begin{table}[!h]
\caption{Range of $m$ allowed by $Y_B$ constraint for different $m_0$ values with set 1, 3, 5 of $\{p,~q,~\alpha,~\beta\}$}
\label{range_m2}
\vspace{5mm}
\begin{adjustwidth}{-2cm}{}
\begin{tabular}{p{1.8cm}|p{1.8cm}|p{1.8cm}|p{1.8cm}|p{1.8cm}|p{1.8cm}|p{1.8cm}|p{1.8cm}|p{1.8cm}|}
\cline{2-9} 
  & \multicolumn{6}{c|}{set 1}   & set 3 & set 5 \\
\hline
\multicolumn{1}{|c|}{$m_0\times10^9$ (GeV)}& $1.881$  & $1.907$ & $1.933$ & $1.959$ & $1.984$ & $2.009$ & $1.717$ & $1.683$ \\
\hline
\multicolumn{1}{|c|}{$\frac{m}{10^7}$ (GeV)}& $4.51-4.62$ & $4.44-4.55$ & $4.37-4.48$ & $4.31-4.42$ & $4.25-4.36$& $4.19-4.30$& $1.47-1.51$&$1.80-1.85$\\
\hline 
\end{tabular}
\end{adjustwidth}
\end{table}
\begin{table}[!h]
\caption{Sets of parameters allowed by both oscillation data and baryon asymmetry bound for case(ii) of 
symmetry breaking with $ \epsilon=-0.004$ (In fully flavored regime)} \label{full_case2}
\begin{center}
\begin{tabular}{p{1cm}|p{1cm}|p{1cm}|p{1.5cm}|p{1.5cm}|p{1.5cm}|p{2.5cm}|}
\cline{2-7}
 & \multicolumn{6}{c|}{parameters}\\
\hline
\multicolumn{1}{ |c| }{sets} &  $p$ & $q$ & $\alpha$ (deg.) & $\beta$ (deg.)&  $m_0\times10^9$ (GeV)& $\frac{m}{10^7}$ (GeV)\\ 
\cline{1-7}
\multicolumn{1}{ |c| }{1 } &  $0.97$  & $1.05$  & $-125$  & $116$ &  $1.881$  & $4.51-4.62$  \\
\multicolumn{1}{ |c| }{ } &    &   &   &  &  $1.907$  &   $4.44-4.55$ \\ 
\multicolumn{1}{ |c| }{ } &    &   &   &  & $1.933$   &   $4.37-4.48$ \\ 
\multicolumn{1}{ |c| }{ } &    &   &   &  & $1.959$   &   $4.31-4.42$ \\ 
\multicolumn{1}{ |c| }{ } &    &   &   &  & $1.984$   &   $4.25-4.36$\\ 
\multicolumn{1}{ |c| }{ } &    &   &   &  & $2.009$   &  $4.19-4.30$ \\
\hline
%
\multicolumn{1}{ |c| }{3 } &  $0.89$  & $0.95$  & $-121$  & $113.5$ &  $1.717$  & $1.47-1.51$   \\
\hline
\multicolumn{1}{ |c| }{5 } &  $0.91$  & $0.91$  & $-117.5$  & $116$ &  $1.683$  &   $1.80-1.85$ \\
\hline
\end{tabular}
\end{center}
\end{table}
\paragraph{}
In {\bf $\tau$-flavored} regime we encounter exactly the same consequences as in the previous case 
(case(i) of symmetry breaking), i.e the final value of baryon asymmetry ($Y_B$) produced for 
any value of right handed neutrino mass (in the range $10^9<m({\rm GeV})<10^{12}$) with any of the 
$16$ value of the set  $\{p,~q,~\alpha,~\beta,~m_0\}$ is far beyond the the experimental upper bound.
\paragraph{}
In {\bf Unflavored} regime the required CP asymmetry parameters ($\varepsilon_i=\sum\limits_{\alpha}\varepsilon^\alpha_i$) for cyclic symmetric $m_D$ and
$M_R=diag(m,~m+\epsilon,~m)$ comes out to be
$\varepsilon_2=0$ and $\varepsilon_1=-\varepsilon_3$ (detailed calculation shown in appendix \ref{a2}). 
Therefore using the same argument as the previous
case (case(i) of symmetry breaking) it can be shown that at the starting point of iteration 
$\frac{d Y_\Delta}{dz}=0$ and hence baryogenesis through leptogenesis is not possible in this regime.

We conclude the analysis of case(ii) with a remark that imposition of baryon asymmetry bound together
with neutrino oscillation data constrains the right handed neutrino mass scale as well as selects only
those parameters belonging to set 1, 3, 5. Hence the value of  $\delta_{\rm CP}$ can only be
positive as: $44.45^\circ$ for set 1, $16.56^\circ$ for set 3 and $14.27^\circ$ for set 5.

\subsection{Numerical analysis for case(iii) of symmetry breaking }
This variant of symmetry breaking arises due to incorporation of the breaking parameter in the \textquoteleft$11$\textquoteright~
element of $M_R$ and the neutrino physics observable (mass squared differences and mixing angles) are calculated with the 
$m_\nu$ matrix given in eq.(\ref{mnu3}). In this case too the lowest allowed value of breaking parameter ($\epsilon$) is $-0.004$
for which the parameter space allowed by oscillation data is presented in Table \ref{case3}.
\begin{table}[!h]
\caption{Sets of parameters allowed by oscillation data for case(iii) of symmetry breaking with $ \epsilon=-0.004$} \label{case3}
\begin{center}
\begin{tabular}{p{1cm}|p{1cm}|p{1cm}|p{1.5cm}|p{1.5cm}|p{1.5cm}|p{2cm}|}
\cline{2-7}
 & \multicolumn{6}{c|}{parameters}\\
\hline
\multicolumn{1}{ |c| }{sets} &  $p$ & $q$ & $\alpha$ (deg.) & $\beta$ (deg.)&  $m_0\times10^9$ (GeV)& $\frac{m}{10^7}$ (GeV)\\ 
\cline{1-7}
\multicolumn{1}{ |c| }{1 } &  $0.91$  & $0.91$  & $-116$  & $117.5$ &  $1.695$  &   \\
\multicolumn{1}{ |c| }{ } &    &   &   &  &  $1.717$  &  unrestricted  \\ 
\multicolumn{1}{ |c| }{ } &    &   &   &  & $1.739$   &  for all \\ 
\multicolumn{1}{ |c| }{ } &    &   &   &  & $1.761$   &  $m_0$ values \\ 
\hline
\multicolumn{1}{ |c| }{2 } &  $0.91$  & $0.91$  & $116$  & $-117.5$ &  $1.695$  &    \\
\multicolumn{1}{ |c| }{ } &    &   &   &  &  $1.717$  &  unrestricted \\
\multicolumn{1}{ |c| }{ } &    &   &   &  &  $1.739$  &  for all  \\
\multicolumn{1}{ |c| }{ } &    &   &   &  &   $1.761$ &  $m_0$ values\\
\hline
\multicolumn{1}{ |c| }{3 } &  $0.95$  & $0.89$  & $-113.5$  & $121$ &  $1.717$  &  unrestricted  \\
\hline
\multicolumn{1}{ |c| }{4 } &  $0.95$  & $0.89$  & $121$  & $-113.5$ &  $1.717$  &  unrestricted  \\
\hline
\multicolumn{1}{ |c| }{5 } &  $1.05$  & $0.97$  & $-116$  & $125$ &  $1.881$  &    \\
\multicolumn{1}{ |c| }{ } &    &   &   &  &  $1.907$  &  unrestricted  \\ 
\multicolumn{1}{ |c| }{ } &    &   &   &  & $1.933$   &  for all \\ 
\multicolumn{1}{ |c| }{ } &    &   &   &  & $1.959$   &  $m_0$ values \\ 
\multicolumn{1}{ |c| }{ } &    &   &   &  & $1.984$   &   \\
\multicolumn{1}{ |c| }{ } &    &   &   &  & $2.009$   &   \\
\hline
\multicolumn{1}{ |c| }{6 } &  $1.05$  & $0.97$  & $116$  & $-125$ &  $1.881$  &   \\
\multicolumn{1}{ |c| }{ } &    &   &   &  &  $1.907$  &  unrestricted  \\ 
\multicolumn{1}{ |c| }{ } &    &   &   &  & $1.933$   &  for all \\ 
\multicolumn{1}{ |c| }{ } &    &   &   &  & $1.959$   &  $m_0$ values \\ 
\multicolumn{1}{ |c| }{ } &    &   &   &  & $1.984$   &   \\
\multicolumn{1}{ |c| }{ } &    &   &   &  & $2.009$   &   \\
\hline
\end{tabular}
\end{center}
\end{table}
 For the case (iii) 
normal hierarchy is preferred and $\theta_{23}$ is  selected in the 1st octant ($40.90^\circ-41.92^\circ$).
The sign of $\alpha$ and $\beta$  remain unsettled between the following sets: (1 and 2), (3 and 4), (5 and 6). 
These in effect again produce sign ambiguity in Dirac CP phase: 
 $\delta_{\rm CP}=14.41^\circ$ for the set 1 with ($\alpha=-116^\circ$, $\beta=117.5^\circ$)  and
 $\delta_{\rm CP}=-14.41^\circ$ for the set 2 with ($\alpha=116^\circ$, $\beta=-117.5^\circ$),
$\delta_{\rm CP}=16.36^\circ$ for the set 3 with ($\alpha=-113.5^\circ$, $\beta=121^\circ$)  and
 $\delta_{\rm CP}=-16.36^\circ$ for the set 4 with ($\alpha=113.5^\circ$, $\beta=-121^\circ$),
$\delta_{\rm CP}=44.41^\circ$ for the set 5 with ($\alpha=-116^\circ$, $\beta=125^\circ$)  and
 $\delta_{\rm CP}=-44.41^\circ$ for the set 6 with ($\alpha=116^\circ$, $\beta=-125^\circ$).

\begin{center}
\begin{figure}[!h]
\includegraphics[width=6.5cm,height=5.5cm,angle=0]{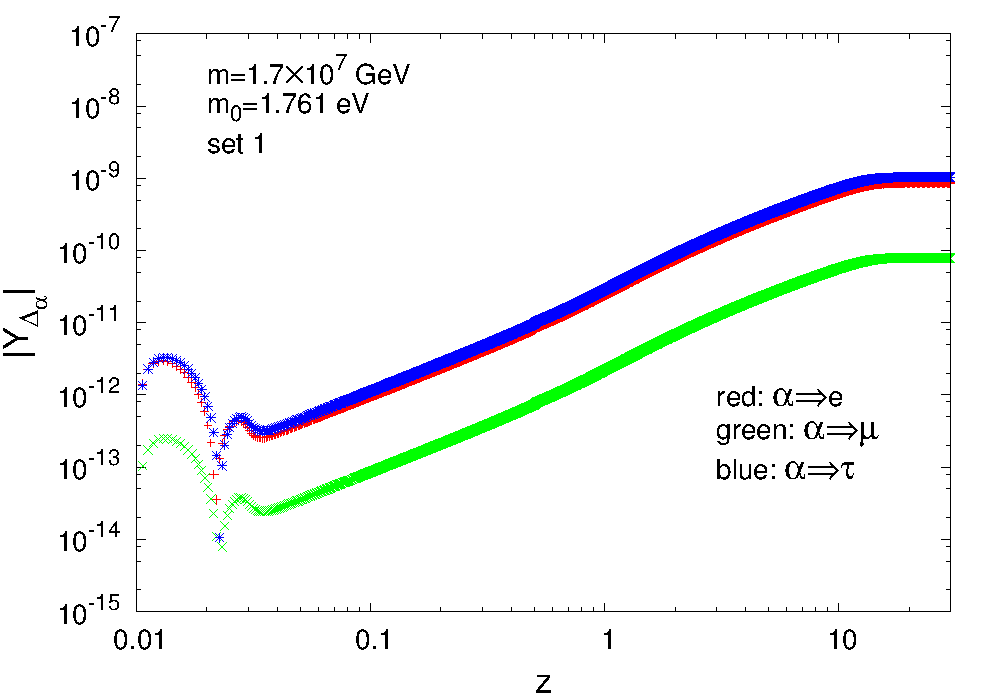}
\hspace{1.0cm}
\includegraphics[width=6.5cm,height=5.5cm,angle=0]{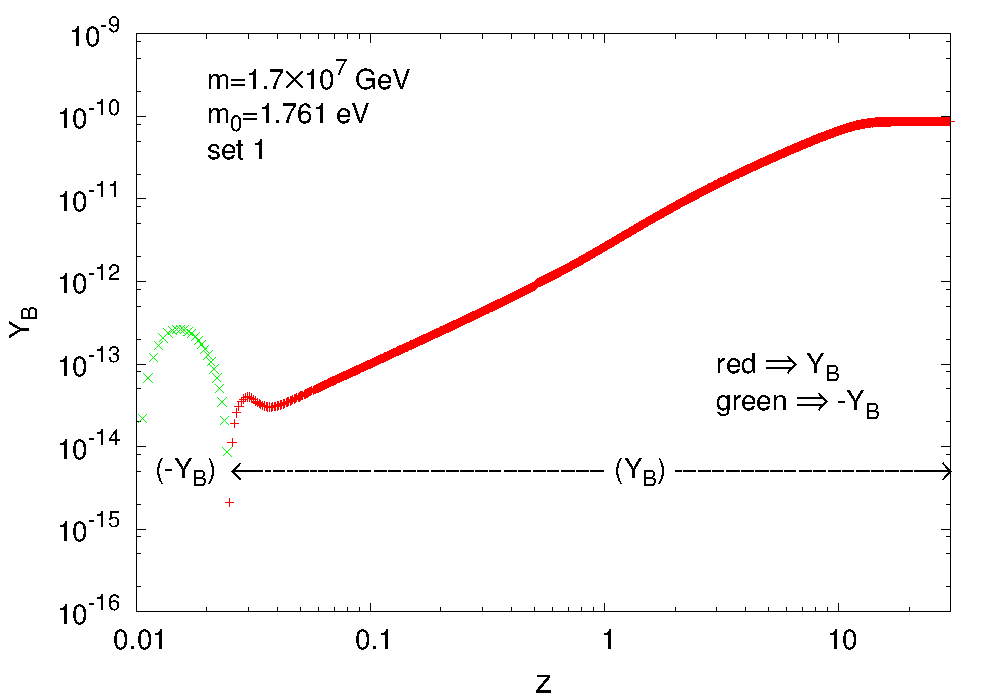}
\caption{(colour online) Plot of flavor asymmetries($Y_{\Delta_\alpha}$) (left), baryon asymmetry($Y_B$) (right)
with $z$ for a definite value of $m$ and $m_0$. In $Y_B~vs~z$ plot $Y_B$ freezes at a value $8.68\times10^{-11}$.}
\label{etl3}
\end{figure}
 \end{center}

\begin{figure}[!h]
\includegraphics[width=4cm,height=4cm,angle=0]{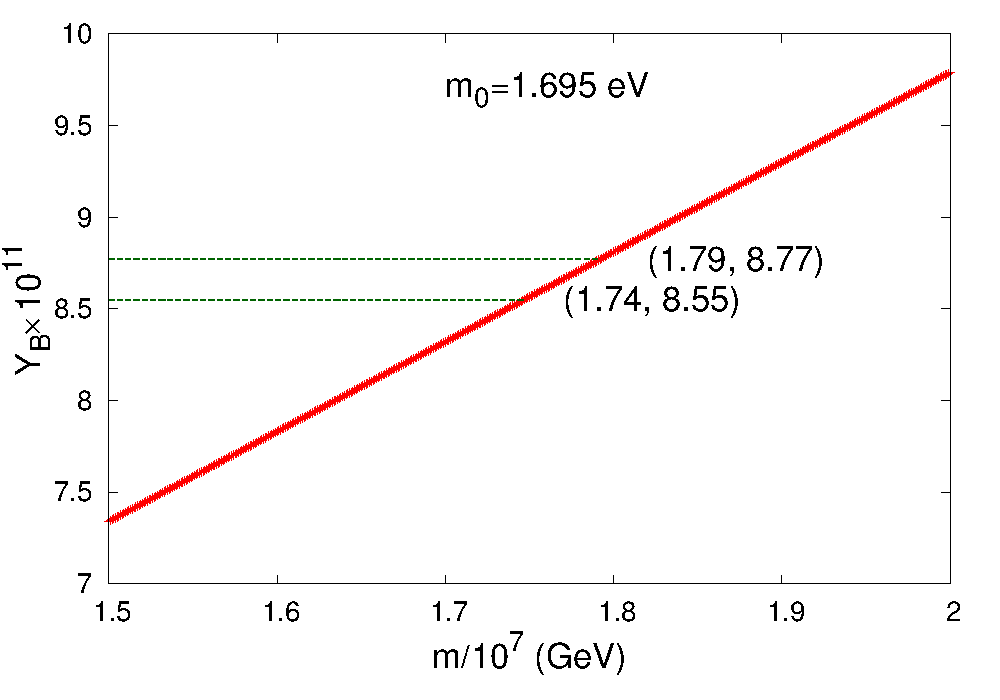}
\includegraphics[width=4cm,height=4cm,angle=0]{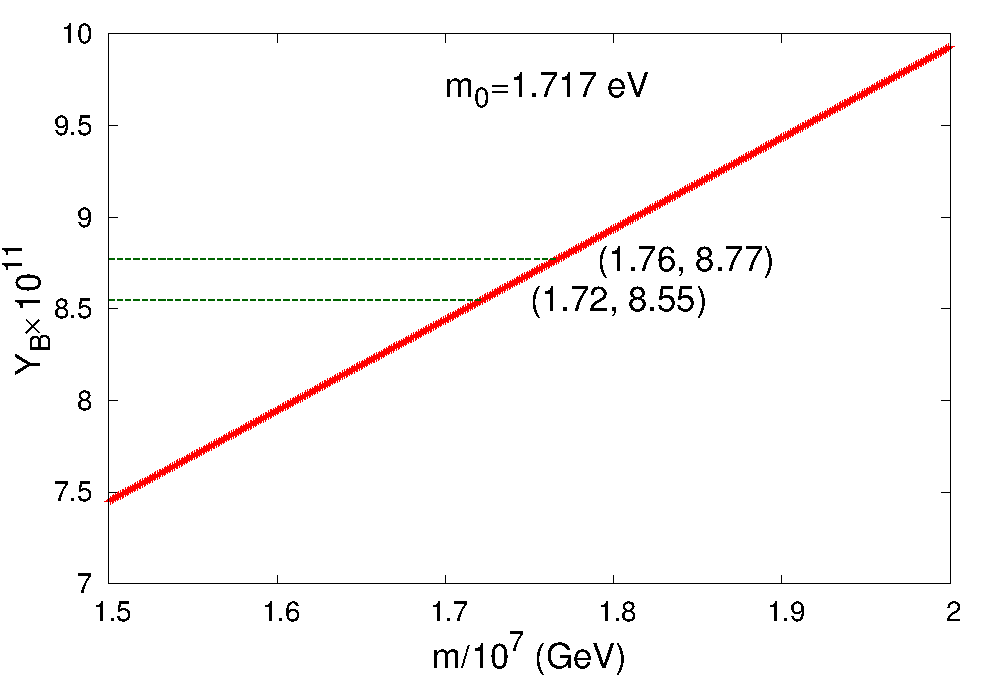}
\includegraphics[width=4cm,height=4cm,angle=0]{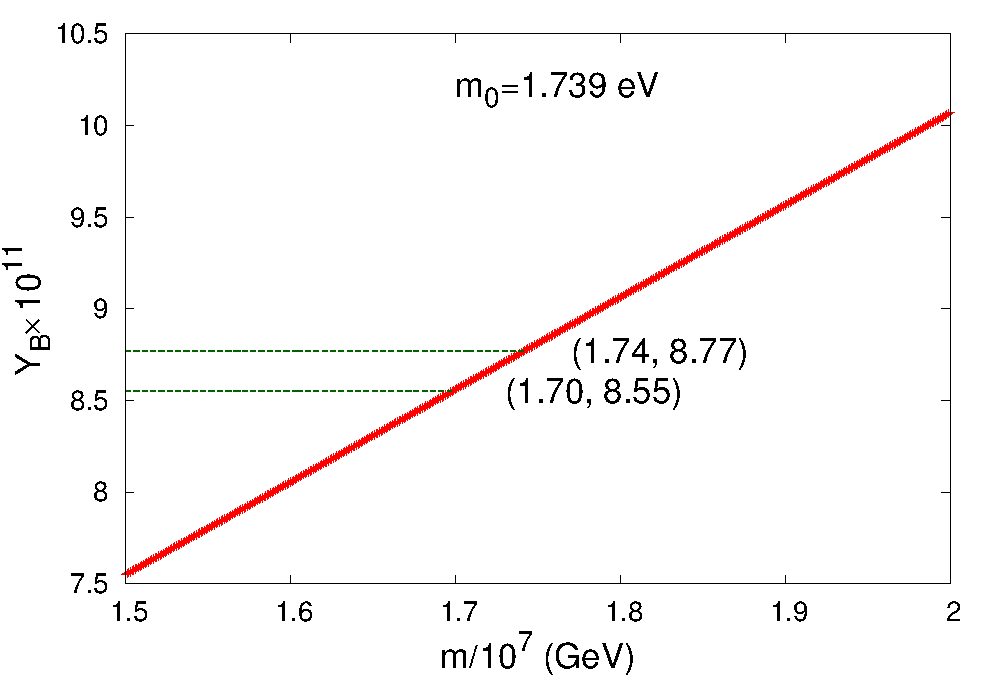}
\includegraphics[width=4cm,height=4cm,angle=0]{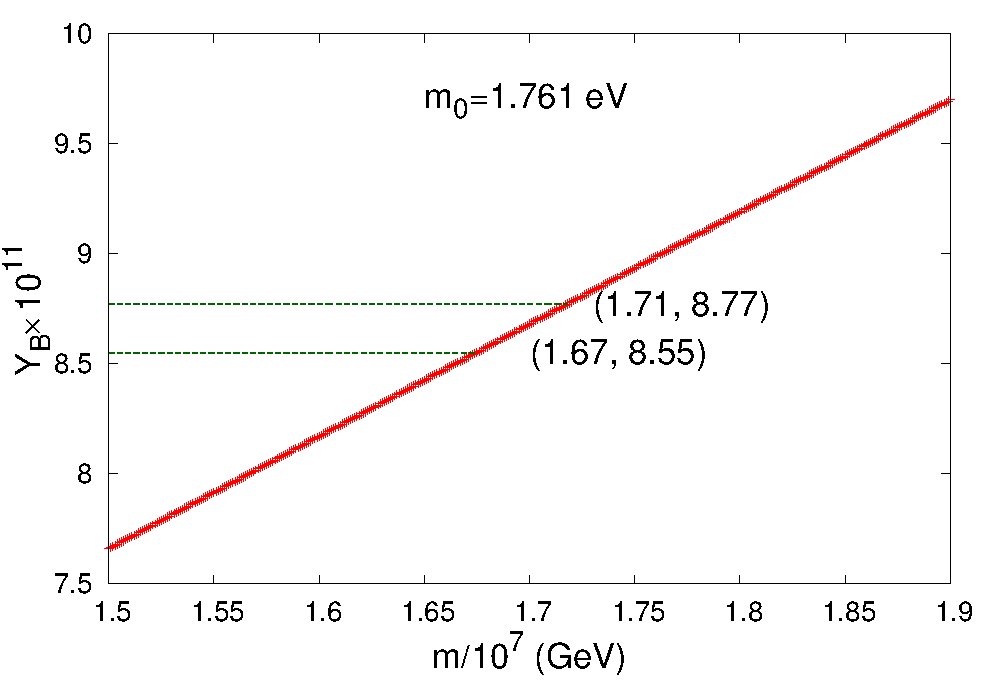}
\caption{(colour online) Plot of Final $Y_B$ with $m$ for different values of $m_0$ with set 1 of $\{p,~q,~\alpha,~\beta\}$}
\label{etab_s1}
\end{figure}

\begin{figure}[!h]
\begin{center}
\includegraphics[width=5cm,height=5cm,angle=0]{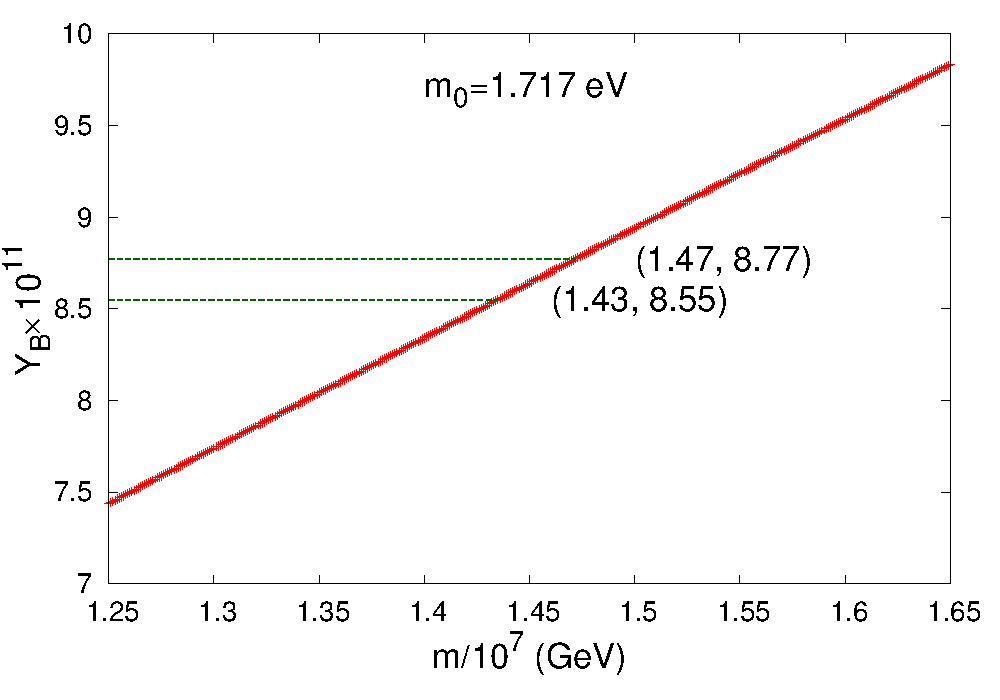}
\caption{(colour online) Plot of Final $Y_B$ with $m$ for different values of $m_0$ with set 3 of $\{p,~q,~\alpha,~\beta\}$}
\label{etab_s2}
\end{center} 
\end{figure}

\begin{figure}[!h]
\begin{center}
\includegraphics[width=5cm,height=5cm,angle=0]{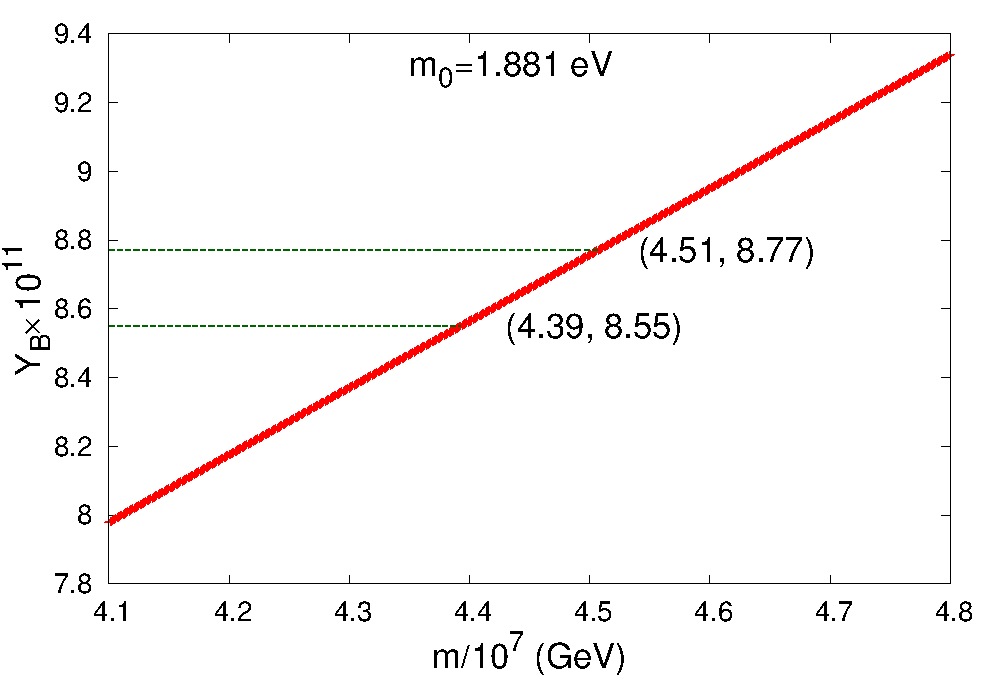}
\includegraphics[width=5cm,height=5cm,angle=0]{2yb_m7.png}
\includegraphics[width=5cm,height=5cm,angle=0]{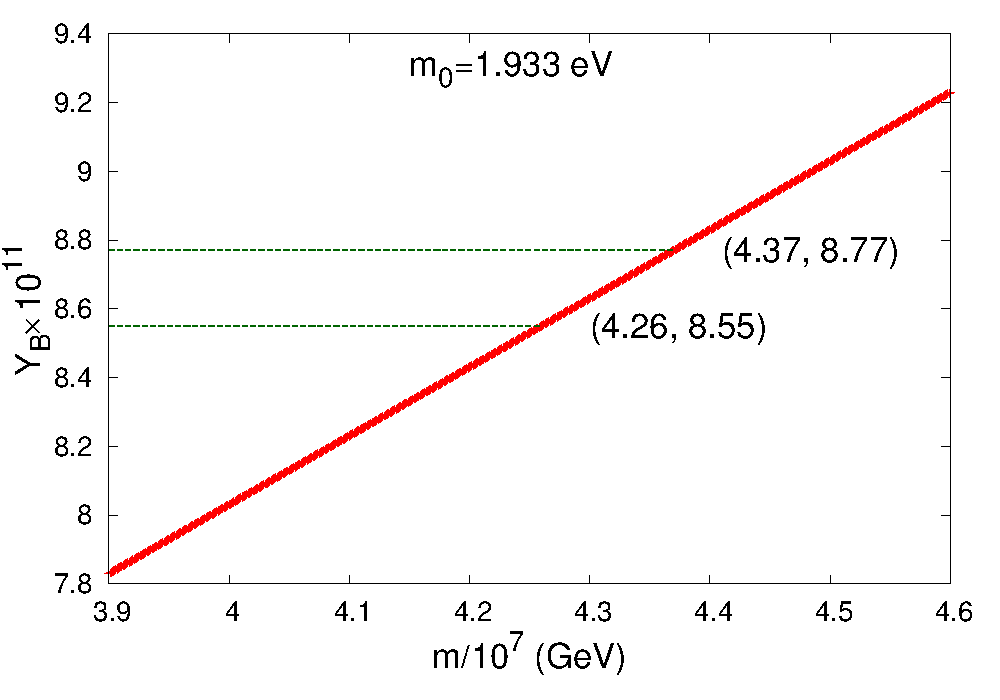}\\
\includegraphics[width=5cm,height=5cm,angle=0]{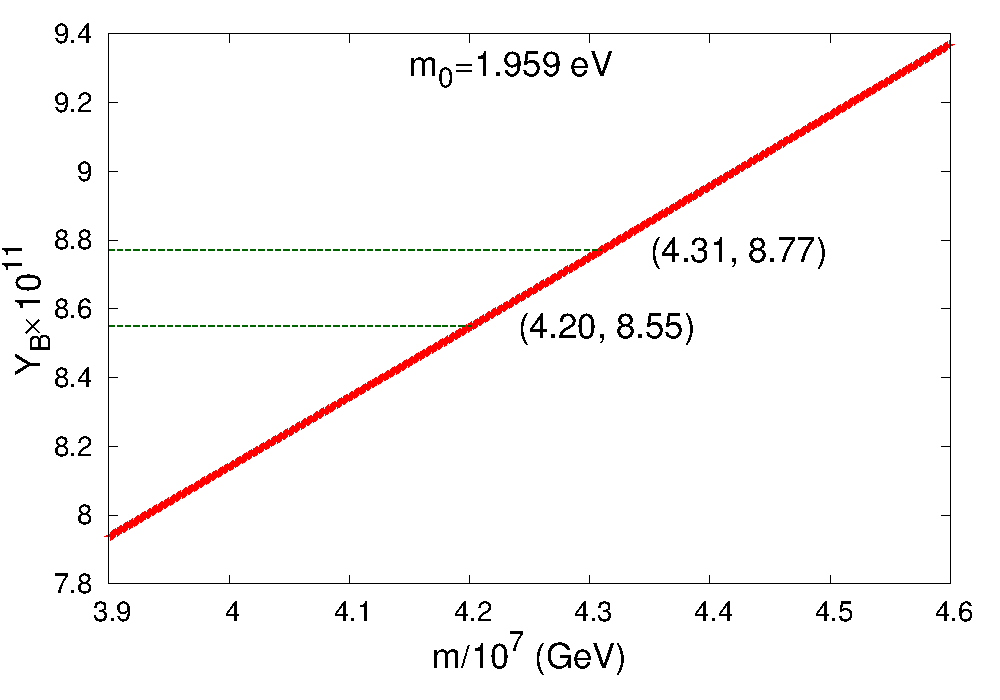}
\includegraphics[width=5cm,height=5cm,angle=0]{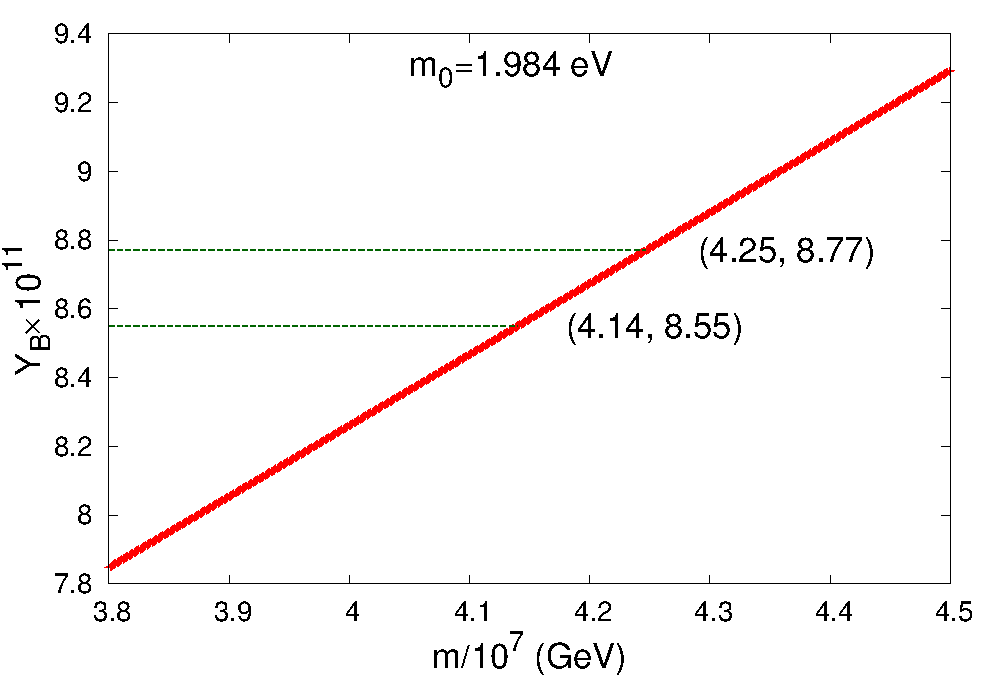}
\includegraphics[width=5cm,height=5cm,angle=0]{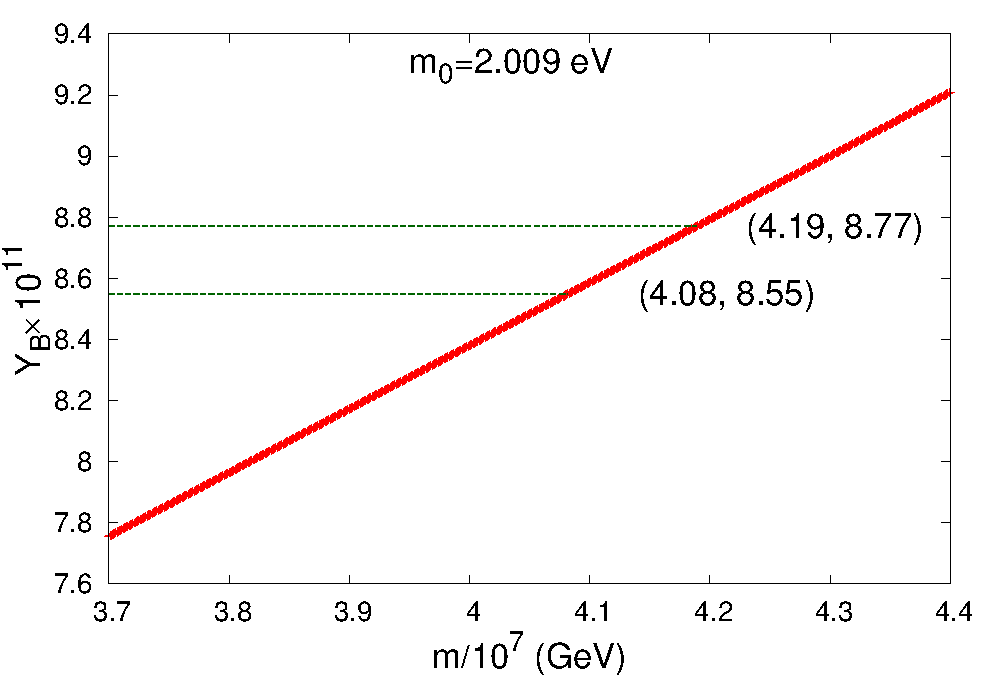}
\caption{(colour online) Plot of Final $Y_B$ with $m$ for different values of $m_0$ with set 5 of $\{p,~q,~\alpha,~\beta\}$}
\label{etab_s3}  
\end{center}
\end{figure}

To solve the sign ambiguity
of the phases and to restrict the right handed neutrino mass scale $m$  for this case also,
we proceed to calculate the baryon asymmetry with these 22 values\footnote{22=4(set 1) +4(set 2) +1(set 3) +1(set 4) +6(set 5) +6(set 6).} 
of the set of parameters $\{p,q,\alpha,\beta,m_0\}$. Following exactly the same procedure as done in the 
previous cases we calculate the final value of baryon asymmetry for all three sub categories of leptogenesis
namely {\bf fully flavored}, {\bf $\tau$-flavored} and {\bf unflavored}.
\paragraph{}
The result obtained in the {\bf fully flavored case} is analogous to that we have got in case(i) and case(ii). In this case
set 2, 4 and 6 generate constant negative value of $Y_B$ at high $z$ and thus those sets are discarded.
It is found that set 1, 3 and 5 are able to produce a $Y_B$ that freezes to a positive value at low temperature.
The final value of baryon asymmetry parameter produced by them for different values of the right handed
neutrino mass $m$ is shown in the Table \ref{final_bary3}. As an example we choose the case 
$(m=1.7\times10^7~{\rm GeV},~m_0=1.761\times10^{-9}~{\rm GeV})$ of set 1 and show the evolution of
flavor asymmetries ($Y_{\Delta_\alpha}$) and baryon asymmetry ($Y_B$) with $z$ in Fig. \ref{etl3}. 
The sign of different flavor asymmetry parameters for this chosen example is given in Table \ref{asy3} and one important
outcome of numerical analysis in this case is $|Y_{\Delta_e}|<|Y_{\Delta_\mu}+Y_{\Delta_\tau}|$.
The plots of $Y_B$ vs $m$ to get a bound on right handed
neutrino mass are given in Figures \ref{etab_s1}, \ref{etab_s2}, \ref{etab_s3} and the bound obtained from those
plots are tabulated clearly in Table \ref{range_m3}.
Finally, the fully constrained parameter space for case(iii) of symmetry breaking is presented in Table \ref{full_case3}.
\begin{table}[!h]
\caption{Final value of baryon asymmetry for set 1, 3, 5 of $\{ p,~q,~\alpha,~\beta \}$ ($m$ and $m_0$ are in GeV)} \label{final_bary3}
\begin{center}
\begin{tabular}{p{2cm}|p{1cm}|p{1cm}|p{1cm}|p{1cm}|p{1cm}|p{1cm}|p{1cm}|p{1cm}|p{1cm}|}
\cline{2-9}
 & \multicolumn{8}{c|}{\bf $Y_B\times10^{11}$ }\\
\cline{1-9}
\multicolumn{1}{ |c| }{\backslashbox{$m_0\times10^9$}{$\frac{m}{10^7}$}} & $1.3$  & $1.5$ & $1.7$ & $2.0$ & $4.0$ & $4.3$ & $4.6$ & $5.0$ \\ \cline{1-10} 
\multicolumn{1}{ |c| }{ $1.695$ } & $6.36$ & $7.34$ & $8.32$ & $9.79$ & $19.58$  & $21.04$ & $22.51$ & $24.47$ &\\
\cline{1-9}
\multicolumn{1}{ |c| }{$1.717$ } & $6.45$ & $7.45$ & $8.44$ & $9.93$ & $19.86$ & $21.35$ & $22.84$ & $24.83$  &set 1\\
\cline{1-9}
\multicolumn{1}{ |c| }{ $1.739$} & $6.55$ & $7.55$ & $8.56$ & $10.07$ & $20.15$ & $21.66$ & $23.17$ & $25.19$ & \\
\cline{1-9}
\multicolumn{1}{ |c| }{$1.761$ } & $6.64$ & $7.66$ & $8.68$ & $10.21$ & $20.43$ & $21.96$ & $23.50$ & $25.54$ &\\
\hline 
\multicolumn{1}{ |c| }{$1.717$ } & $7.74$ & $8.94$ & $10.13$ & $11.92$ & $23.84$ & $25.62$ & $27.41$ & $29.80$ &set 3\\
\hline
\multicolumn{1}{ |c| }{$1.881$ } & $2.53$ & $2.92$ & $3.31$ & $3.89$ & $7.78$ & $8.37$ & $8.95$ & $9.73$ &\\
\cline{1-9}
\multicolumn{1}{ |c| }{$1.907$ } & $2.57$ & $2.96$ & $3.36$ & $3.95$ & $7.91$ & $8.50$ & $9.09$ & $9.88$ &set 5\\
\cline{1-9}
\multicolumn{1}{ |c| }{$1.933$ } & $2.60$ & $3.01$ & $3.41$ & $4.01$ & $8.03$ & $8.63$ & $9.23$ & $10.03$ &\\
\cline{1-9}
\multicolumn{1}{ |c| }{$1.959$ } & $2.64$ & $3.05$ & $3.46$ & $4.07$ & $8.14$ & $8.75$ & $9.37$ & $10.18$ &\\
\cline{1-9}
\multicolumn{1}{ |c| }{$1.984$ } & $2.68$ & $3.10$ & $3.51$ & $4.13$ & $8.26$ & $8.88$ & $9.50$ & $10.33$ &\\
\cline{1-9}
\multicolumn{1}{ |c| }{$2.009$ } & $2.72$ & $3.14$ & $3.56$ & $4.19$ & $8.38$ & $9.00$ & $9.63$ & $10.47$ &\\
\hline
\end{tabular}
\end{center}
\end{table}
\begin{table}[!h]
\caption{signs of different asymmetries at different $z$ values} \label{asy3}
\begin{center}
\begin{tabular}{p{2cm}|p{.8cm}p{.8cm}p{.8cm}p{.8cm}|p{.8cm}p{.8cm}p{.8cm}p{.8cm}|}
\cline{2-9}
 &\multicolumn{4}{c|}{$z=0.01\rightarrow 0.02$}& \multicolumn{4}{c|}{$z>0.02$}\\
\hline
\multicolumn{1}{ |c| }{}& $Y_{\Delta_e}$  & $Y_{\Delta_\mu}$ & $Y_{\Delta_\tau}$ & $Y_B$  & $Y_{\Delta_e}$  & $Y_{\Delta_\mu}$ & $Y_{\Delta_\tau}$ & $Y_B$ \\ \cline{1-9}
\multicolumn{1}{ |c| }{set 6 } & +ve &-ve &-ve & -ve & -ve  & +ve & +ve & +ve \\
\hline
\end{tabular}
\end{center}
\end{table}

\begin{table}[!h]
\caption{Range of $m$ allowed by $Y_B$ constraint for different $m_0$ values with set 1, 3, 5 of $\{p,~q,~\alpha,~\beta\}$}
\label{range_m3}
\vspace{5mm}
\begin{adjustwidth}{-1.5cm}{}
\begin{tabular}{p{1.8cm}|p{1.2cm}|p{1.1cm}|p{1.1cm}|p{1.1cm}|p{1.1cm}|p{1.1cm}|p{1.1cm}|p{1.1cm}|p{1.1cm}|p{1.1cm}|p{1.1cm}|}
\cline{2-12} 
  & \multicolumn{4}{c|}{set 1}   & set 3 & \multicolumn{6}{c|}{set 5} \\
\hline
\multicolumn{1}{|c|}{$m_0\times10^9$ (GeV)}& $1.695$  & $1.717$ & $1.739$ & $1.761$ & $1.717$ & $1.881$ & $1.907$ & $1.933$ &$1.959$ & $1.984$& $2.009$\\
\hline
\multicolumn{1}{|c|}{$\frac{m}{10^7}$ (GeV)}& $1.74-$ & $1.72-$ & $1.70-$ & $1.67-$ & $1.43-$& $4.39-$& $4.32-$ &$4.26-$&$4.20$ & $4.14-$&$4.08-$\\
\multicolumn{1}{|c|}{}                      & $1.79$  & $1.76$ & $1.74$ & $1.71$ & $1.47$    & $4.51$& $4.43$ & $4.37$& $4.31$& $4.25$&$4.19$\\
\hline 
\end{tabular}
\end{adjustwidth}
\end{table}
\begin{table}[!h]
\caption{Sets of parameters allowed by both oscillation data and baryon asymmetry bound for case(iii) of 
symmetry breaking with $ \epsilon=-0.004$ (In fully flavored regime)} \label{full_case3}
\begin{center}
\begin{tabular}{p{1cm}|p{1cm}|p{1cm}|p{1.5cm}|p{1.5cm}|p{1.5cm}|p{3cm}|}
\cline{2-7}
 & \multicolumn{6}{c|}{parameters}\\
\hline
\multicolumn{1}{ |c| }{sets} &  $p$ & $q$ & $\alpha$ (deg.) & $\beta$ (deg.)&  $m_0\times10^9$ (GeV)& $\frac{m}{10^7}$ (GeV)\\ 
\cline{1-7}
\multicolumn{1}{ |c| }{1 } &  $0.91$  & $0.91$  & $-116$  & $117.5$ &  $1.695$  &  $1.74-1.79$  \\
\multicolumn{1}{ |c| }{ } &    &   &   &  &  $1.717$  &  $1.72-1.76$ \\
\multicolumn{1}{ |c| }{ } &    &   &   &  &  $1.739$  &   $1.70-1.74$ \\
\multicolumn{1}{ |c| }{ } &    &   &   &  &   $1.761$ &  $1.67-1.71$\\
\hline
\multicolumn{1}{ |c| }{3 } &  $0.95$  & $0.89$  & $-121$  & $113.5$ &  $1.717$  & $1.43-1.47$    \\
\hline
\multicolumn{1}{ |c| }{5 } &  $1.05$  & $0.97$  & $-116$  & $125$ &  $1.881$  &  $4.39-4.51$ \\
\multicolumn{1}{ |c| }{ } &    &   &   &  &  $1.907$  &   $4.32-4.43$ \\ 
\multicolumn{1}{ |c| }{ } &    &   &   &  & $1.933$   &   $4.26-4.37$\\ 
\multicolumn{1}{ |c| }{ } &    &   &   &  & $1.959$   &   $4.20-4.31$\\ 
\multicolumn{1}{ |c| }{ } &    &   &   &  & $1.984$   &   $4.14-4.25$\\
\multicolumn{1}{ |c| }{ } &    &   &   &  & $2.009$   &   $4.08-4.19$\\
\hline
\end{tabular}
\end{center}
\end{table}
\paragraph{}
Results of {\bf $\tau$-flavored} and {\bf unflavored} case are same as that of case(i) and (ii).
In {\bf $\tau$-flavored} regime set 1, 3 and 5 generate a positive value of baryon asymmetry 
which is far beyond the experimental upper limit, whereas in the {\bf unflavored} regime\footnote{
CP asymmetry parameters ($\varepsilon_i=\sum\limits_{\alpha}\varepsilon^\alpha_i$) for cyclic symmetric $m_D$ and
$M_R=diag(m+\epsilon,~m,~m)$ comes out to be $\varepsilon_1=0$ and $\varepsilon_2=-\varepsilon_3$ 
(detailed calculation shown in appendix \ref{a3}). Therefore using the same argument as the previous
case (case(i) of symmetry breaking) it can be shown that at the starting point of iteration 
$\frac{d Y_\Delta}{dz}=0$ and hence generation of baryon asymmetry is not possible. }
generation of baryon asymmetry is not at all possible.

Again the baryon asymmetry bound along with oscillation data admits only those parameters parameters 
belonging to set 1, 3, 5 as well as constrains mass scale of the right handed neutrino. 
Hence in this case also  $\delta_{\rm CP}$ can have only positive values: $14.41^\circ$ for set 1, 
$16.36^\circ$ for set 3 and $41.41^\circ$ for set 5.


\vspace{4mm}
\paragraph{}
It is evident from the numerical analysis of the three cases 
(case(i), (ii) and (iii)) that baryon asymmetry in the allowed range can
only be generated in fully flavored regime. For $\tau$-flavored regime all three cases 
produce excess baryon asymmetry. The phases $\alpha,~\beta$ are restricted by light neutrino
data and they are not closer to $0$ or $\pi$ also. This $\tau$-flavored regime is closer 
to the resonant enhancement region
and phase suppression does not occur. So, it goes beyond experimental range. For unflavored regime in all three cases
our breaking mechanism of cyclic symmetry and summation over flavor produce null contribution to the lepton asymmetry and hence baryon asymmetry
although the resonant enhancement of $\varepsilon^\alpha_i$'s occur in this regime. 
\section{Summary}\label{summ}
We consider an $SU(2)_L\times U(1)_Y$ model with three right chiral neutrinos invoking type-I seesaw mechanism and cyclic symmetry
in the neutrino sector. Since, the symmetry invariant model generates two fold degeneracy in the light neutrino mass, the 
model forbids to determine three mixing angles in an unique way as well as generates vanishing value of one mass squared
difference. A possible way to get rid of those shortcomings is due to the breaking of the cyclic symmetry imposed.
Symmetry breaking is incorporated in a minimal way through one small breaking parameter ($\epsilon$).
~Armed with such modifications, we apply the most general diagonalization method to find out mass eigenvalues, 
mixing angles and Dirac CP phase. First we restrict the parameter space by fixing the neutrino oscillation experimental data.
The first level of restriction is done for three different cases of symmetry breaking, viz 
$(i)~M_R=diag\{m,~m,~m(1+\epsilon)\},~(ii)~M_R=diag\{m,~m(1+\epsilon),~m\}~{\rm and}~(iii)~M_R=diag\{m(1+\epsilon),~m,~m\} $.
We have seen that normal hierarchy of light neutrino masses is preferred for all three cases. The obtained parameter space prefer
$\theta_{23}$ to be in the first octant ($\simeq 37^\circ$) for case (i), 2nd octant ($48^\circ-49^\circ$ ) for case (ii)
and 1st octant ($40^\circ-41^\circ$ )for case (iii).
Furthermore the sign ambiguity in $\alpha$ and $\beta$ produce both positive and negative values of $\delta_{\rm CP}$
prior to the application of baryon asymmetry bound.
\paragraph{}
Next we investigate explicitly the effect of quasi degeneracy of right handed neutrino masses
in enhancement of CP asymmetry. Among the allowed values of the set of Lagrangian parameters we choose only those 
sets corresponding to the lowest value of breaking parameter. 
Only those values of the parameter sets are used in calculation of leptogenesis.
The phenomena of leptogenesis is studied in three different energy
regimes $\{(m({\rm GeV})<10^9),(10^9<m({\rm GeV})<10^{12}),(m({\rm GeV})>10^{12})\}$
where lepton flavors are fully distinguishable, partly distinguishable or indistinguishable respectively.
Calculation of lepton asymmetry in these regimes are carried out 
thereafter solving the detailed set of Boltzmann equations. 
These lepton asymmetries are then converted to baryon asymmetry using suitable formulas.
\paragraph{}
Notable outcomes of our numerical analysis are:
\begin{itemize}
\item{ Only fully flavored leptogenesis 
is able to produce baryon asymmetry in the observed range. Unflavored leptogenesis is unable to generate 
any asymmetry in all the cases. $\tau$-flavored leptogenesis although analytically allowed however numerical
estimation shows that value of produced asymmetry is far beyond the present experimental limit.}

\item{Using the cut on $Y_B$ we
 have obtained a bound on right handed heavy neutrino mass $(1.43-4.62)\times 10^7$ GeV 
(considering all the cases (case(i), case(ii) and case(iii))
of fully flavored regime) which were unconstrained even after  the restriction by neutrino oscillation data.}

\item{Dirac CP phase takes only positive value in the range $14^\circ-45^\circ$ after imposition of baryon asymmetry 
bound (considering all three cases of symmetry breaking).}
\end{itemize}
\appendix
\section{Appendix}
\subsection{CP asymmetry parameters in unflavored regime with case(i) of symmetry breaking ($M_R=diag\{m,~m~,m(1+\epsilon)\}$)}\label{a1}
The flavor summed CP asymmetry parameter relevant in this regime is given by 
$\varepsilon_i=\sum\limits_{\alpha}\varepsilon^\alpha_i$:
\begin{equation}
\varepsilon_i=\frac{1}{4 \pi v^2 H_{ii}} \sum\limits_{j \neq i} Im\{H_{ij}^2\} g(x_{ij})
\end{equation}
where $g(x_{ij})=f(x_{ij})+\frac{\sqrt{x_{ij}}(1-x_{ij})}{(1-x_{ij})^2+\frac{H_{jj}^2}{16\pi^2 v^4}}$ and 
$x_{ij}=\frac{m_{N_j}^2}{m_{N_i}^2}$.\\\\
In this breaking scheme we have 
\begin{equation}
x_{12}=1,~x_{23}=x_{13}=(1+\epsilon)^2~~{\rm and}~~x_{ji}=\frac{1}{x_{ij}}.
\end{equation}
The elements of $H$ matrix (eq.(\ref{capH})) are obtained as
\begin{eqnarray}
&&H_{12}=H_{23}=H_{31}=m m_0 Y =re^{i\theta}\nonumber\\
&&H_{13}=H_{21}=H_{32}=m m_0 Y^\ast =re^{-i\theta}.
\end{eqnarray}
Using these values we get 
\begin{eqnarray}
\varepsilon_1 &=& \frac{1}{4 \pi v^2 H_{11}} [ Im\{ H_{12}^2 \} g(x_{12})+ Im\{ H_{13}^2 \} g(x_{13})] \nonumber\\
            &=& \frac{1}{4 \pi v^2 H_{11}}[r^2 \sin 2\theta g(1) -r^2 \sin 2\theta g((1+\epsilon)^2)],
\end{eqnarray}
\begin{eqnarray}
 \varepsilon_2 &=& \frac{1}{4 \pi v^2 H_{22}} [ Im\{ H_{21}^2 \} g(x_{21})+ Im\{ H_{23}^2 \} g(x_{23})] \nonumber\\
            &=& \frac{1}{4 \pi v^2 H_{22}}[-r^2 \sin 2\theta g(1) +r^2 \sin 2\theta g((1+\epsilon)^2)].
\end{eqnarray}
Therefore it is clear that $\varepsilon_1=-\varepsilon_2$ since in our model $H_{11}=H_{22}=H_{33}$. 
Finally
\begin{eqnarray}
 \varepsilon_3 &=& \frac{1}{4 \pi v^2 H_{33}} [ Im\{ H_{31}^2 \} g(x_{31})+ Im\{ H_{32}^2 \} g(x_{32})] \nonumber\\
            &=& \frac{1}{4 \pi v^2 H_{22}}[r^2 \sin 2\theta g\{\frac{1}{(1+\epsilon)^2}\} -r^2 \sin 2\theta g\{\frac{1}{(1+\epsilon)^2}\}]\nonumber\\
            &=& 0
\end{eqnarray}
\subsection{CP asymmetry parameters in unflavored regime with case(ii) of symmetry breaking ($M_R=diag\{m,~m(1+\epsilon),~m\}$)}\label{a2}
In this breaking scheme we have 
\begin{equation}
x_{12}=\frac{1}{x_{23}}=(1+\epsilon)^2,~x_{13}=1~~{\rm and}~~x_{ji}=\frac{1}{x_{ij}}.
\end{equation}
The elements of $H$ matrix (eq.(\ref{capH})) are obtained as
\begin{eqnarray}
&&H_{12}=H_{23}=H_{31}=m m_0 Y =re^{i\theta}\nonumber\\
&&H_{13}=H_{21}=H_{32}=m m_0 Y^\ast =re^{-i\theta}.
\end{eqnarray}
Using these values we get 
\begin{eqnarray}
\varepsilon_1 &=& \frac{1}{4 \pi v^2 H_{11}} [ Im\{ H_{12}^2 \} g(x_{12})+ Im\{ H_{13}^2 \} g(x_{13})] \nonumber\\
            &=& \frac{1}{4 \pi v^2 H_{11}}[r^2 \sin 2\theta g((1+\epsilon)^2) -r^2 \sin 2\theta g(1) ],
\end{eqnarray}
\begin{eqnarray}
 \varepsilon_3 &=& \frac{1}{4 \pi v^2 H_{33}} [ Im\{ H_{31}^2 \} g(x_{31})+ Im\{ H_{32}^2 \} g(x_{32})] \nonumber\\
            &=& \frac{1}{4 \pi v^2 H_{22}}[r^2 \sin 2\theta g(1) -r^2 \sin 2\theta g((1+\epsilon)^2)]
\end{eqnarray}
i.e we get $\varepsilon_1=-\varepsilon_3$ and
\begin{eqnarray}
 \varepsilon_2 &=& \frac{1}{4 \pi v^2 H_{22}} [ Im\{ H_{21}^2 \} g(x_{21})+ Im\{ H_{23}^2 \} g(x_{23})] \nonumber\\
            &=& \frac{1}{4 \pi v^2 H_{22}}[-r^2 \sin 2\theta g\{\frac{1}{(1+\epsilon)^2}\}   +r^2 \sin 2\theta g\{\frac{1}{(1+\epsilon)^2}\} ]\nonumber\\
            &=&0
\end{eqnarray}
\subsection{CP asymmetry parameters in unflavored regime with case(iii) of symmetry breaking ($M_R=diag\{m(1+\epsilon),~m,~m\}$)}\label{a3}
In this breaking scheme we have 
\begin{equation}
x_{12}=x_{13}=\frac{1}{(1+\epsilon)^2},~x_{23}=1~~{\rm and}~~x_{ji}=\frac{1}{x_{ij}}.
\end{equation}
The elements of $H$ matrix (eq.(\ref{capH})) are obtained as
\begin{eqnarray}
&&H_{12}=H_{23}=H_{31}=m m_0 Y =re^{i\theta}\nonumber\\
&&H_{13}=H_{21}=H_{32}=m m_0 Y^\ast =re^{-i\theta}.
\end{eqnarray}
Using these values we get 
\begin{eqnarray}
\varepsilon_1 &=& \frac{1}{4 \pi v^2 H_{11}} [ Im\{ H_{12}^2 \} g(x_{12})+ Im\{ H_{13}^2 \} g(x_{13})] \nonumber\\
            &=& \frac{1}{4 \pi v^2 H_{11}}[r^2 \sin 2\theta g\{(1+\epsilon)^{-2}\} -r^2 \sin 2\theta g\{(1+\epsilon)^{-2}\} ]\nonumber\\
            &=&0
\end{eqnarray}
\begin{eqnarray}
 \varepsilon_2 &=& \frac{1}{4 \pi v^2 H_{22}} [ Im\{ H_{21}^2 \} g(x_{21})+ Im\{ H_{23}^2 \} g(x_{23})] \nonumber\\
            &=& \frac{1}{4 \pi v^2 H_{22}}[-r^2 \sin 2\theta g\{(1+\epsilon)^2\}   +r^2 \sin 2\theta g\{\ 1 \} ]
\end{eqnarray}
\begin{eqnarray}
 \varepsilon_3 &=& \frac{1}{4 \pi v^2 H_{33}} [ Im\{ H_{31}^2 \} g(x_{31})+ Im\{ H_{32}^2 \} g(x_{32})] \nonumber\\
            &=& \frac{1}{4 \pi v^2 H_{22}}[r^2 \sin 2\theta g\{(1+\epsilon)^2\} -r^2 \sin 2\theta g\{1\}]
\end{eqnarray}
i.e we get $\varepsilon_2=-\varepsilon_3$ .


\newpage
\begin{thebibliography}{10}
\bibitem{Rubakov:1996vz} 
  V.~A.~Rubakov and M.~E.~Shaposhnikov,
  Usp.\ Fiz.\ Nauk {\bf 166}, 493 (1996)
  [Phys.\ Usp.\  {\bf 39}, 461 (1996)]
  [hep-ph/9603208].
\bibitem{Trodden:1998ym} 
  M.~Trodden,
  Rev.\ Mod.\ Phys.\  {\bf 71}, 1463 (1999)
  [hep-ph/9803479].
\bibitem{Riotto:1998bt} 
  A.~Riotto,
  [hep-ph/9807454].

\bibitem{Cline:2006ts} 
  J.~M.~Cline,
  [hep-ph/0609145].
\bibitem{Dine:2003ax} 
  M.~Dine and A.~Kusenko,
  Rev.\ Mod.\ Phys.\  {\bf 76}, 1 (2003)
  [hep-ph/0303065].
\bibitem{sak} A.D. Sakharov, Zh. Eksp. Teor. Fiz. Pis'ma {\bf 5}, 32 (1967); JETP Lett. {\bf 91B}, 24 (1967).
\bibitem{Dev:2015uca} 
  P.~S.~B.~Dev and R.~N.~Mohapatra,
Phys.\ Rev.\ D {\bf 92}, no. 1, 016007 (2015)
  [arXiv:1504.07196 [hep-ph]].
\bibitem{Ade:2015xua} 
  P.~A.~R.~Ade {\it et al.} [Planck Collaboration],
  [arXiv:1502.01589 [astro-ph.CO]].
\bibitem{Ade:2013zuv} 
  P.~A.~R.~Ade {\it et al.}  [Planck Collaboration],
Astron.\ Astrophys.\  {\bf 571}, A16 (2014)
  [arXiv:1303.5076 [astro-ph.CO]].
\bibitem{one}
M.~Fukugita and T.~Yanagida, 
Phys.\ Lett.\  B {\bf 174} (1986) 45.

\bibitem{one2}
A.~Riotto and M.~Trodden, 
Ann.\ Rev.\ Nucl.\ Part.\ Sci.\  {\bf 49} (1999) 35 [arXiv:hep-ph/9901362].

\bibitem{khlopov}
M.~Yu.~Khlopov, {\it Cosmoparticle\,\, Physics, World Scientific, Singapore} 
(1999).
\bibitem{Davidson:2008bu} 
  S.~Davidson, E.~Nardi and Y.~Nir,
  Phys.\ Rept.\  {\bf 466}, 105 (2008)
  [arXiv:0802.2962 [hep-ph]].
\bibitem{Buchmuller:2000as} 
  W.~Buchmuller and M.~Plumacher,
  Int.\ J.\ Mod.\ Phys.\ A {\bf 15}, 5047 (2000)
  [hep-ph/0007176].
\bibitem{Bernreuther:2002uj} 
  W.~Bernreuther,
  Lect.\ Notes Phys.\  {\bf 591}, 237 (2002)
  [hep-ph/0205279].
\bibitem{Forero:2014bxa} 
  D.~V.~Forero, M.~Tortola and J.~W.~F.~Valle,
Phys.\ Rev.\ D {\bf 90}, no. 9, 093006 (2014)
  [arXiv:1405.7540 [hep-ph]].
\bibitem{GonzalezGarcia:2012sz} 
  M.~C.~Gonzalez-Garcia, M.~Maltoni, J.~Salvado and T.~Schwetz,
  JHEP {\bf 1212}, 123 (2012)
  [arXiv:1209.3023 [hep-ph]].
\bibitem{Tortola:2012te} 
  D.~V.~Forero, M.~Tortola and J.~W.~F.~Valle,
  Phys.\ Rev.\ {\bf D 86} (2012) 073012
  [arXiv:1205.4018 [hep-ph]].

\bibitem{Adhikary:2013bma} 
  B.~Adhikary, M.~chakraborty and A.~Ghosal,
 JHEP {\bf 1310}, 043 (2013)
  Erratum: [JHEP {\bf 1409}, 180 (2014)]
 [arXiv:1307.0988 [hep-ph]].
\bibitem{Koide:2000zi} 
  Y.~Koide,
  [hep-ph/0005137].

\bibitem{Damanik:2007cs} 
  A.~Damanik, M.~Satriawan, P.~Anggraita, A.~Hermanto and Muslim,
  J.\ Theor.\ Comput.\ Stud.\  {\bf 8}, (2008) 0102
  [arXiv:0710.1742 [hep-ph]].
\bibitem{Damanik:2010rv} 
  A.~Damanik,
  [arXiv:1004.1457 [hep-ph]].
\bibitem{Samanta:2015oqa} 
  R.~Samanta and A.~Ghosal,
  [arXiv:1507.02582 [hep-ph]].
\bibitem{Covi:1996wh} 
  L.~Covi, E.~Roulet and F.~Vissani,
  Phys.\ Lett.\ B {\bf 384}, 169 (1996)
  [hep-ph/9605319].

\bibitem{Pilaftsis:2003gt} 
  A.~Pilaftsis and T.~E.~J.~Underwood,
  Nucl.\ Phys.\ B {\bf 692}, 303 (2004)
  [hep-ph/0309342].
\bibitem{Pilaftsis:1997jf} 
  A.~Pilaftsis,
  Phys.\ Rev.\ D {\bf 56}, 5431 (1997)
  [hep-ph/9707235].
\bibitem{Adhikary:2010fa} 
  B.~Adhikary, A.~Ghosal and P.~Roy,
  JCAP {\bf 1101}, 025 (2011)
  [arXiv:1009.2635 [hep-ph]].
\bibitem{MAL} M.~A.~Luty, Phys.\ Rev.\ {\bf D45} (1992) 455.

\bibitem{kolb_turner}
 E.~W.~Kolb and M.~S.~Turner, {\it The Early Universe} 
(Addison-Wesley, Redwood City, CA, U.S.A) (1990).

\bibitem{Adhikary:2006rf} 
  B.~Adhikary,
  Phys.\ Rev.\ D {\bf 74}, 033002 (2006)
  [hep-ph/0604009].
\bibitem{k3}
 A.~Abada, S.~Davidson, A.~Ibarra, F.~-X.~Josse-Michaux, M.~Losada and A.~Riotto,
  JHEP {\bf 0609} (2006) 010
  [hep-ph/0605281].
\bibitem{antush} 
  S.~Antusch, S.~F.~King and A.~Riotto,
  JCAP {\bf 0611}, 011 (2006)
  [hep-ph/0609038].
\bibitem{harvey}
J A.~Harvey, M S.~Turner,
Phys.\ Rev.\ D {\bf 42}, 3344 (1990)

 \bibitem{Giusarma:2013pmn}
  E.~Giusarma, R.~de Putter, S.~Ho and O.~Mena,
 Phys.\ Rev.\ D {\bf 88}, no. 6, 063515 (2013)
  arXiv:1306.5544 [astro-ph.CO].
\bibitem{Bennett:2012zja} 
  C.~L.~Bennett {\it et al.}  [WMAP Collaboration],
  Astrophys.\ J.\ Suppl.\  {\bf 208}, 20 (2013)
  [arXiv:1212.5225 [astro-ph.CO]].
\bibitem{Aihara:2011sj} 
  H.~Aihara {\it et al.}  [SDSS Collaboration],
  Astrophys.\ J.\ Suppl.\  {\bf 193}, 29 (2011)
  [Erratum-ibid.\  {\bf 195}, 26 (2011)]
  [arXiv:1101.1559 [astro-ph.IM]].
\bibitem{Auger:2012ar}
  M.~Auger {\it et al.}  [EXO Collaboration],
  Phys.\ Rev.\ Lett.\  {\bf 109} (2012) 032505
  [arXiv:1205.5608 [hep-ex]].
\bibitem{Giuliani:2010zz} 
  A.~Giuliani,
  Acta Phys.\ Polon.\  {\bf B 41} (2010) 1447.
\bibitem{Rodejohann:2012xd} 
  W.~Rodejohann,
  J.\ Phys.\ {\bf G 39} (2012) 124008 
  [arXiv:1206.2560 [hep-ph]].
\bibitem{Ghosal:2001ep} 
  A.~Ghosal, Y.~Koide and H.~Fusaoka,
  Phys.\ Rev.\ D {\bf 64}, 053012 (2001)
  [hep-ph/0104104].
\end{thebibliography}
\end{document}